\documentclass[a4paper,12pt]{article}
\pdfoutput=1
\usepackage{epsfig,graphicx,xcolor,amsbsy,amssymb,latexsym,amsfonts,amsmath,,tcolorbox,setspace}
\usepackage{pstricks}
\usepackage{color}
\usepackage{soul}
\usepackage[numbers,square,comma, compress]{natbib}
\usepackage{placeins}
\usepackage{wrapfig,framed,caption}
\usepackage[makeroom]{cancel}

\usepackage{wasysym}

\usepackage{multirow}

\definecolor{shadecolor}{gray}{0.925}

\usepackage{eurosym}
\usepackage[a4paper]{geometry}
\usepackage{hyperref}
\usepackage{graphicx}
\usepackage{tikz}
\usetikzlibrary{decorations.pathreplacing,decorations.markings,snakes}
\usetikzlibrary{decorations.pathmorphing}
\usetikzlibrary{patterns}
\usetikzlibrary{calc}
\usepackage{xcolor}
\usepackage{amsmath,amssymb,amsfonts,pstricks,setspace}
\usepackage[enableskew]{youngtab}

\numberwithin{equation}{section}

\usepackage{wrapfig}

\newcommand{\bea}{\begin{eqnarray}\displaystyle}
\newcommand{\eea}{\end{eqnarray}}

\title{
{\bf Renormalisation Group Methods  \\ for \\ Effective Epidemiological Models  }\\[40pt]
}

\author{\large \textsc{Stefan~Hohenegger\footnote{\tt s.hohenegger@ipnl.in2p3.fr}}~,~\,and\,~\textsc{Francesco Sannino\footnote{\tt sannino@qtc.sdu.dk}}}

\begin{document}

\maketitle
\thispagestyle{empty}
\begin{center}
\renewcommand{\thefootnote}{\fnsymbol{footnote}}\vspace{-0.5cm}
${}^{\footnotemark[1]}$ Univ Lyon, Univ Claude Bernard Lyon 1, CNRS/IN2P3, IP2I Lyon, UMR 5822, F-69622, Villeurbanne, France\\
${}^{\footnotemark[2]}\,{}^{\footnotemark[3]}$ Dept. of Physics E. Pancini, Università di Napoli Federico II, via Cintia, 80126 Napoli, Italy\\[0.2cm]
 ${}^{\footnotemark[2]}$ Scuola Superiore Meridionale, Largo S. Marcellino, 10, 80138 Napoli NA, Italy\\
${}^{\footnotemark[2]}$ INFN sezione di Napoli, via Cintia, 80126 Napoli, Italy\\
${}^{\footnotemark[2]}$ Quantum  Theory Center ($\hbar$QTC) \& D-IAS, Southern Denmark Univ., Campusvej 55, 5230 Odense M, Denmark\\[2.5cm]
\end{center}

\begin{abstract}
Epidemiological models describe the spread of an infectious disease within a population. They capture microscopic details on how the disease is passed on among individuals in various different ways, while making predictions about the state of the entirety of the population. However, the type and structure of the specific model considered typically depend on the size of the population under consideration. To analyse this effect, we study a family of  effective epidemiological models in space and time that are related to each other through scaling transformations. Inspired by a similar treatment of diffusion processes, we interpret the latter as renormalisation group transformations, both at the level of the underlying differential equations and their solutions. We  show that in the large scale limit, the microscopic details of the infection process become irrelevant, safe for a simple real number, which plays the role of the infection rate in a basic compartmental model.
\end{abstract}

\newpage

\tableofcontents

\section{Introduction}
The mathematical modelling of the spread of an infectious diseases through a population has a longstanding history. Indeed, numerous conceptually different approaches have been discussed in the literature (see \emph{e.g.}~\cite{BaileyBook,Becker,DietzSchenzle,Frauenthal,Castillo,Capasso,DaleyGani,DiekmannHeesterbeek,Lauwerier,Dietz,Dietz2,AndersonMayBook,AndersonMay,HethcoteThousand,Cacciapaglia:2021vvu} for reviews). Classifying them in an exhaustive fashion is rather difficult, however, a common distinction sets apart\footnote{A complete separation is indeed difficult, since models can contain elements of both. Furthermore, more recent data- and computer-driven approaches are difficult to attribute to either of the two.}  
\begin{itemize}
\item {\bf stochastic approaches:} the basic processes that govern the spread of the disease (\emph{e.g.} the infection or recovery of individuals) are of a probabilistic nature. Models using this approach are usually discretised (\emph{i.e.} the population consists of discretely many individuals and time is a discrete variable) and are characterised by a set of probabilities (\emph{e.g.} the probability for an infected individual to infect another one in close proximity). 
\item {\bf deterministic approaches:} the spread of the disease is understood as a predictable process in time. The number of individuals that are affected (in various different ways) by the disease are described by differential equations in space and time. Models using this approach are characterised by a set of rates (\emph{e.g.} the rate at which infectious individuals infect others). 
\end{itemize}
The choice of approach largely depends on the concrete question under consideration: on the one hand side, since a stochastic modelling of the (microscopic) processes that lead to the spread of a disease is usually closer to the (biological) reality, deterministic models are typically considered as approximations (see \emph{e.g.} \cite{Kendall,Mollison1977}). Stochastic models are well adapted for (short term) simulations of diseases, but their potentially higher computational cost makes long-term predictions difficult. On the other hand, deterministic models try to capture properties of the disease that arise from the basic microscopic processes. If done correctly, due to the intrinsic nature of these models, they allow predictions of general trends and (long-term) developments of the disease.

Among other things, a crucial aspect for the choice of the approach is the scale at which the time-evolution of the disease is described, both in terms of the size of the population and the time-frame. For example, a model that is well adapted to describe the initial outbreak and spread of Covid-19 in a small town might be much less successful in predicting the cumulative number of infected individuals in a large country at the end of an entire wave. Indeed, even though the basic (biological) microscopic processes that govern the transmission of the disease (as well as the recovery) on an individual level are the same, the dynamics in the two cases can be influenced very differently by other aspects, such as initial conditions (\emph{i.e.} the number and geographic distribution of infected individuals in the beginning), the impact of so-called super-spreaders (\emph{i.e.} individuals with a significantly higher chance of infecting other individuals) and super-spreader events (\emph{i.e.} events during which the disease is transmitted much more than usual), the geographic mobility of the population, \emph{etc.}. Depending on the disease in question, this list can further be extended to include socio-geographic, economical and political factors.

A similar scale dependence occurs in many physical problems (see \emph{e.g.} \cite{Barenblatt,IntermediateAsymptotics}), for example phase transitions in thermodynamical systems, effective descriptions of quantum field theory and particle physics or even quantum aspects of black holes (see \emph{e.g.} \cite{Bonanno:2000ep,Binetti:2022xdi,DAlise:2023hls,DelPiano:2023fiw}). Despite the large number of vastly different examples, a general framework to approach them exists in the form of \emph{renormalisation group} methods \cite{oono1985advances,goldenfeld2018lectures,PhysRevLett.72.76,PhysRevE.49.4502}, which for example in \cite{ChenGoldenfeldOono1995} have been characterised in the following fashion: {\it 'The essence of the renormalisation group method is to extract structurally stable features of a system which are insensitive to details.'} The current paper is dedicated to applying this idea to epidemiology.

Indeed, we consider as a template model the well studied relation between diffusion processes and random walks: for sufficiently large times, the probability (density) to find a random walker in a finite region of a $d$-dimensional (regular) lattice can be approximated by a function of continuous space- and time variables. Furthermore, in a specific continuum limit, this probability density becomes the causal Green's function of the heat equation, which vanishes at spatial infinity. We establish a similar line of reasoning in the case of a simplistic model of epidemiology: the starting point is a stochastic model on a regular lattice, where the lattice sites represent individuals, who may be infected with a communicable disease and the probability for a healthy individual to become infected is modelled as a function of the infectious individuals at neighbouring lattice sites. We approximate this stochastic model by a compartmental model \cite{Kermack:1927} (see also \cite{BaileyBook,Becker,DietzSchenzle,Castillo,Dietz,Dietz2,HethcoteThousand} for reviews), where the number of infectious individuals is given as a function of continuous space- and time-variables that satisfy a first order non-linear differential equation in time \cite{Mollison1977,MollisonVelocities}. The details of how individuals infect one another depending on their spatial position is encoded in a single integrable function $f\in L^1$. We argue that a continuous rescaling of the spatial variable leads to a family of effective models, that describe the average number of infected individuals in a variable region of space. In the limit of the latter becoming infinite, we recover the well known SI-model \cite{Kermack:1927}: in this case, the entire information of the function $f$ is scaled away, except for its $L^1$-norm, which is interpreted as the infection rate. Since this rescaling procedure falls into the definition given in \cite{ChenGoldenfeldOono1995} cited above, we propose this mechanism to describe a renormalisation group method in epidemiology, as advocated previously in \cite{DellaMorte:2020wlc,DellaMorte:2020qry,cacciapaglia2020second,cacciapaglia2020mining,cacciapaglia2020evidence,cacciapaglia2020multiwave,Cacciapaglia:2020mjf,cacciapaglia2020us,Cacciapaglia:2021cjl,Cacciapaglia:2021vvu,GreenPass,MLvariants} based on phenomenological arguments. Indeed, the SI-model found in the limiting case of our family of epidemic models coincides with the simplest version of the so-called \emph{epidemiological renormalisation group} (eRG) equation proposed in \cite{DellaMorte:2020wlc}.

This work focuses on the mathematical and conceptual ideals, thus sacrificing epidemiological realistic features for computational simplicity: for example, our family of models (following \cite{Mollison1977}) does not allow recovery of infectious individuals and infections are instantaneous in time (\emph{i.e.} we do not allow for an incubation time of the disease). Furthermore, countermeasures of the population (vaccinations, non-pharmaceutical interventions, \emph{etc.}) are neglected as are details of the mobility or the structure of the population. We plan to discuss more sophisticated models in the future.

Concretely, this work is organised as follows: In Section~\ref{Sect:SimpleExamples} we first consider the simple example of diffusion processes, which allows us to introduce various concepts related to scaling, namely coarse graining of degrees of freedom which leads to families of self-similar models. In Section~\ref{Sect:SimpleEpidemicSI} we shall apply these concepts to a simple model of epidemiology~\cite{Mollison1977,MollisonVelocities}, leading to the definition of a family of related models, whose fixed point is the SI-model \cite{Kermack:1927}. Finally, Section~\ref{Sect:Conclusions} contains our conclusions and an outlook for future work. This paper is supplemented by two appendices, which contain additional mathematical and conceptual definitions as well as computational details that were deemed to lengthy for the main body of the text.

\section{Diffusion: Random Walks and Heat Equation}\label{Sect:SimpleExamples}

Before discussing epidemiological models that describe the spread of a disease throughout a population, we discuss a simple system in which the concept of scaling can be made evident, namely a simple diffusion process in $d$-dimensions. This examples showcases a number of features that we shall implement into the description of epidemiological processes (concretely into the 'simple' epidemic model described in \cite{MollisonVelocities}) in the subsequent Section.

We first present a stochastic approach to the phenomenon of diffusion in the form of random walks on a (regular) lattice. This description can be approximated by deterministic equations, which in the limit of very small lattice spacing become the $d$-dimensional heat equations. While many of the computations and conceptual ideas are well-known in the literature (see \emph{e.g.} the textbook \cite{Itzykson:1989sx} for a pedagogical presentation that we shall follow in this Section), this example allows us to showcase the deterministic approximation of a stochastic process in a controllable setting: we shall discuss the approximation both at the level of the probabilities and the equation governing the dynamics. 
\subsection{Stochastic Processes and Difference Equations}
\begin{figure}[htbp]
\begin{center}
\includegraphics[width=7.5cm]{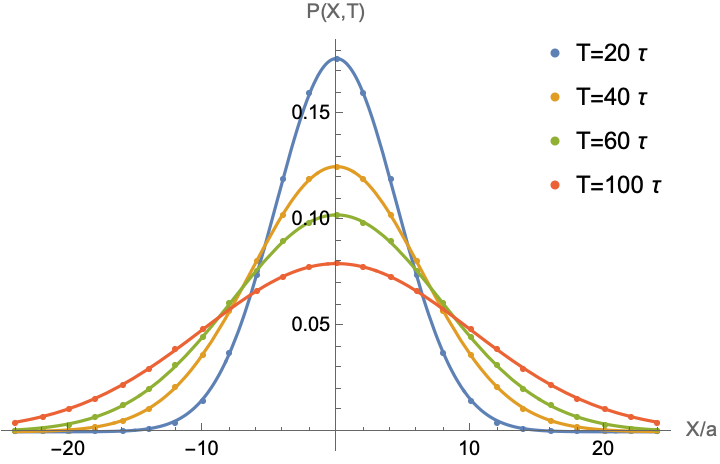}\hspace{1cm}\includegraphics[width=7.5cm]{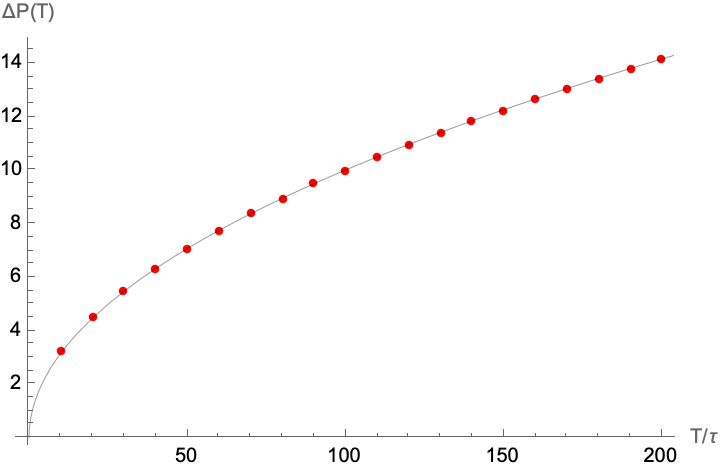}
\end{center}
\caption{Random walks in one dimension ($d=1$): The left panel shows the probability (\ref{PFourierInt}) for different times $T$. The dots represent the values of the integral (\ref{PFourierInt}) (for fixed $a$), while the solid lines are interpolations in the form of Gaussian functions. The right panel shows the width of the later, which is interpolated by $\Delta P(T)=\sqrt{T/\tau}$.}
\label{Fig:RandomWalks}
\end{figure}

Let $\Gamma$ be a $d$-dimensional infinite, hypercubic lattice in Euclidean space
\begin{align}
\Gamma=\{\vec{X}\in\mathbb{R}^d|\vec{X}=\sum_{i=1}^d n_i\,\vec{E}_i\text{ with }n_i\in\mathbb{Z}\}\,,\label{DefHyperLattice}
\end{align}
that is generated by $d$ orthogonal vectors $(\vec{E}_1,\ldots,\vec{E}_d)$ 
\begin{align}
&\vec{E}_i\cdot\vec{E}_j=a^2\,\delta_{ij}\,,&&\forall i,j=1,\ldots,d\,.\label{OrthogonalityBase}
\end{align}
Every lattice site has $2d$ neighbours (with lattice spacing $a\in\mathbb{R}_+$) such that the coordination number is $q=2d$.  We consider a walker, who at regular time intervals $\tau\in\mathbb{R}_+$ takes a step in a random direction by moving to one of the neighbouring lattice sites. We assume that the probability to move to each of the $2d$ neighbouring sites is equal and independent of previous movements or the current position. We denote by $P(\vec{X}_1-\vec{X}_0,T_1-T_0)$ the probability for the walker to be at the time $T_1$ at the lattice site $\vec{X}_1$ when they were at lattice site $\vec{X}_0$ at time $T_0$. Due to translational invariance, $P$ only depends on the differences $\vec{X}=\vec{X}_1-\vec{X}_0$ and $T=T_1-T_0$. Furthermore, the probability satisfies the following recursion relation in time
\begin{align}
P(\vec{X},T+\tau)=\frac{1}{2d}\sum_{i=1}^d\left[P(\vec{X}+\vec{E}_i,T)+P(\vec{X}-\vec{E}_i,T)\right]\,,\label{RecursionRelation}
\end{align}
and is normalised to satisfy
\begin{align}
&P(\vec{X},0)=\delta_{\vec{X},0}\,,&&\text{and}&&\sum_{\vec{X}\in\Gamma}P(\vec{X},T)=1\,.\label{CondsRelRandom}
\end{align}
Upon defining the Fourier transform $\widetilde{P}(\vec{K},T)$ of $P(\vec{X},T)$ with respect to the spatial coordinates
\begin{align}
&P(\vec{X},T)=\frac{1}{(2 \pi) ^d}\int_{[-\pi,\pi]^d} d^d K\, e^{\frac{i\vec{K}\cdot\vec{X}}{a^2}}\,\widetilde{P}(\vec{K},T)\,,&&\text{with}&&\int_{[-\pi,\pi]^d} d^d K = \prod_{i=1}^d \int_{-\pi}^\pi dK_i\,,\label{PFourier}
\end{align}
the recursion relation (\ref{RecursionRelation}) and the initial conditions (\ref{CondsRelRandom}) take the form
\begin{align}
\widetilde{P}(\vec{K},T+\tau)=\frac{1}{d}\,\widetilde{P}(\vec{K},T)\sum_{j=1}^d\cos\left(\frac{\vec{K} \cdot \vec{E}_j}{a^2}\right)\,,&&\text{and}&&\widetilde{P}(\vec{K},0)=1\,.\label{RecursionFourier}
\end{align}
They can be solved to give the following integral representation of the probability
\begin{align}
P(\vec{X},T)
&=\frac{a^d}{(2 \pi) ^d}\int_{\left[-\frac{\pi}{a},\frac{\pi}{a}\right]^d} d^d k\, e^{\frac{i\vec{k}\cdot\vec{X}}{a}}\,\left(\frac{1}{d}\sum_{j=1}^d\cos\left(\vec{k} \cdot \vec{e}_j\right)\right)^{\frac{T}{\tau}}\,,\label{PFourierInt}
\end{align}
where we have introduced the following orthonormal set of vectors in $\mathbb{R}^d$ 
\begin{align}
&\vec{e}_i=\frac{1}{a}\,\vec{E}_i\in\mathbb{R}^d\,,&&\text{with} &&\vec{e}_i\cdot\vec{e}_j=\delta_{ij}\,,\label{OrthonormalBasis}
\end{align}
and the new integration variables $k_i=\frac{K_i}{a}$ with
\begin{align}
\vec{k}=\sum_{i=1}^d k_i\,\vec{E}_i=a\,\sum_{i=1}^d k_i\,\vec{e}_i=\frac{\vec{K}}{a}\,.
\end{align}
A plot of the probabilities (\ref{PFourierInt}) for different times $T$ in one dimension ($d=1$) is shown in Figure~\ref{Fig:RandomWalks}.

\subsection{Deterministic Model and Heat Equation}
As a next step we develop a deterministic approximation of the stochastic process of random walks: to this end, we shall replace the probabilities $P(\vec{X},T)$ by coarse grained quantities that are determined by a differential equation which replaces the recursion relation (\ref{RecursionRelation}).
\subsubsection{Coarse Graining: Continuous Space-Time and Probability Densities}\label{Sect:CoarseGrainingDiffusion}
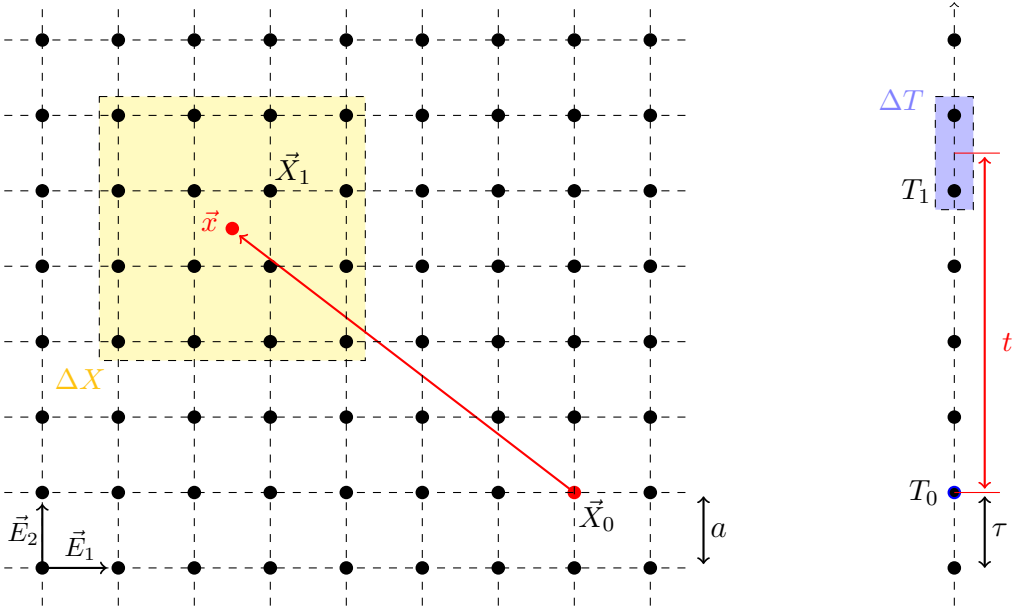
\begin{figure}[htbp]
\begin{center}
\scalebox{1}{\parbox{13.5cm}{\begin{tikzpicture}
\draw[dashed, fill=yellow!30!white] (-3.25,2.25) -- (0.25,2.25) -- (0.25,-1.25) -- (-3.25,-1.25) -- (-3.25,2.25);
%
\draw[thick, fill=black] (-4,3) circle (0.075cm);
\draw[thick, fill=black] (-3,3) circle (0.075cm);
\draw[thick, fill=black] (-2,3) circle (0.075cm);
\draw[thick, fill=black] (-1,3) circle (0.075cm);
\draw[thick, fill=black] (0,3) circle (0.075cm);
\draw[thick, fill=black] (1,3) circle (0.075cm);
\draw[thick, fill=black] (2,3) circle (0.075cm);
\draw[thick, fill=black] (3,3) circle (0.075cm);
\draw[thick, fill=black] (4,3) circle (0.075cm);
\draw[thick, fill=black] (-4,2) circle (0.075cm);
\draw[thick, fill=black] (-3,2) circle (0.075cm);
\draw[thick, fill=black] (-2,2) circle (0.075cm);
\draw[thick, fill=black] (-1,2) circle (0.075cm);
\draw[thick, fill=black] (0,2) circle (0.075cm);
\draw[thick, fill=black] (1,2) circle (0.075cm);
\draw[thick, fill=black] (2,2) circle (0.075cm);
\draw[thick, fill=black] (3,2) circle (0.075cm);
\draw[thick, fill=black] (4,2) circle (0.075cm);
\draw[thick, fill=black] (-4,1) circle (0.075cm);
\draw[thick, fill=black] (-3,1) circle (0.075cm);
\draw[thick, fill=black] (-2,1) circle (0.075cm);
\draw[thick, fill=black] (-1,1) circle (0.075cm);
\draw[thick, fill=black] (0,1) circle (0.075cm);
\draw[thick, fill=black] (1,1) circle (0.075cm);
\draw[thick, fill=black] (2,1) circle (0.075cm);
\draw[thick, fill=black] (3,1) circle (0.075cm);
\draw[thick, fill=black] (4,1) circle (0.075cm);
\draw[thick, fill=black] (-4,0) circle (0.075cm);
\draw[thick, fill=black] (-3,0) circle (0.075cm);
\draw[thick, fill=black] (-2,0) circle (0.075cm);
\draw[thick, fill=black] (-1,0) circle (0.075cm);
\draw[thick, fill=black] (0,0) circle (0.075cm);
\draw[thick, fill=black] (1,0) circle (0.075cm);
\draw[thick, fill=black] (2,0) circle (0.075cm);
\draw[thick, fill=black] (3,0) circle (0.075cm);
\draw[thick, fill=black] (4,0) circle (0.075cm);
\draw[thick, fill=black] (-4,-1) circle (0.075cm);
\draw[thick, fill=black] (-3,-1) circle (0.075cm);
\draw[thick, fill=black] (-2,-1) circle (0.075cm);
\draw[thick, fill=black] (-1,-1) circle (0.075cm);
\draw[thick, fill=black] (0,-1) circle (0.075cm);
\draw[thick, fill=black] (1,-1) circle (0.075cm);
\draw[thick, fill=black] (2,-1) circle (0.075cm);
\draw[thick, fill=black] (3,-1) circle (0.075cm);
\draw[thick, fill=black] (4,-1) circle (0.075cm);
\draw[thick, fill=black] (-4,-2) circle (0.075cm);
\draw[thick, fill=black] (-3,-2) circle (0.075cm);
\draw[thick, fill=black] (-2,-2) circle (0.075cm);
\draw[thick, fill=black] (-1,-2) circle (0.075cm);
\draw[thick, fill=black] (0,-2) circle (0.075cm);
\draw[thick, fill=black] (1,-2) circle (0.075cm);
\draw[thick, fill=black] (2,-2) circle (0.075cm);
\draw[thick, fill=black] (3,-2) circle (0.075cm);
\draw[thick, fill=black] (4,-2) circle (0.075cm);
\draw[thick, fill=black] (-4,-3) circle (0.075cm);
\draw[thick, fill=black] (-3,-3) circle (0.075cm);
\draw[thick, fill=black] (-2,-3) circle (0.075cm);
\draw[thick, fill=black] (-1,-3) circle (0.075cm);
\draw[thick, fill=black] (0,-3) circle (0.075cm);
\draw[thick, fill=black] (1,-3) circle (0.075cm);
\draw[thick, fill=black] (2,-3) circle (0.075cm);
\draw[thick, fill=black,red] (3,-3) circle (0.075cm);
\draw[thick, fill=black] (4,-3) circle (0.075cm);
\draw[thick, fill=black] (-4,-4) circle (0.075cm);
\draw[thick, fill=black] (-3,-4) circle (0.075cm);
\draw[thick, fill=black] (-2,-4) circle (0.075cm);
\draw[thick, fill=black] (-1,-4) circle (0.075cm);
\draw[thick, fill=black] (0,-4) circle (0.075cm);
\draw[thick, fill=black] (1,-4) circle (0.075cm);
\draw[thick, fill=black] (2,-4) circle (0.075cm);
\draw[thick, fill=black] (3,-4) circle (0.075cm);
\draw[thick, fill=black] (4,-4) circle (0.075cm);
\draw[<->,thick] (4.7,-3.95) -- (4.7,-3.05);
\node at (4.9,.-3.5) {$a$};
\node at (3.3,-3.3) {\small $\vec{X}_0$};
\draw[thick, fill=black,red] (-1.5,0.5) circle (0.075cm);
\draw[thick, red,->] (3,-3) -- (-1.41,0.41);
\node[red] at (-1.8,0.6) {\small $\vec{x}$};
\node[yellow!75!red] at (-3.5,-1.5) {\small $\Delta X$};
\node at (-0.7,1.3) {\small $\vec{X}_1$};
%
\draw[dashed] (-4.5,3) -- (4.5,3);
\draw[dashed] (-4.5,2) -- (4.5,2);
\draw[dashed] (-4.5,1) -- (4.5,1);
\draw[dashed] (-4.5,0) -- (4.5,0);
\draw[dashed] (-4.5,-1) -- (4.5,-1);
\draw[dashed] (-4.5,-2) -- (4.5,-2);
\draw[dashed] (-4.5,-3) -- (4.5,-3);
\draw[dashed] (-4.5,-4) -- (4.5,-4);
\draw[dashed] (4,-4.5) -- (4,3.5);
\draw[dashed] (3,-4.5) -- (3,3.5);
\draw[dashed] (2,-4.5) -- (2,3.5);
\draw[dashed] (1,-4.5) -- (1,3.5);
\draw[dashed] (0,-4.5) -- (0,3.5);
\draw[dashed] (-1,-4.5) -- (-1,3.5);
\draw[dashed] (-2,-4.5) -- (-2,3.5);
\draw[dashed] (-3,-4.5) -- (-3,3.5);
\draw[dashed] (-4,-4.5) -- (-4,3.5);
\draw[thick,->] (-4,-4) -- (-3.15,-4);
\node at (-3.5,-3.7) {\footnotesize $\vec{E}_1$};
\draw[thick,->] (-4,-4) -- (-4,-3.15);
\node at (-4.25,-3.5) {\footnotesize $\vec{E}_2$};

\draw[dashed, fill=blue!25!white] (7.75,2.25) -- (8.25,2.25) -- (8.25,0.75) -- (7.75,0.75) -- (7.75,2.25);
\draw[dashed,->] (8,-4.5) -- (8,3.5);
\draw[thick, fill=black] (8,3) circle (0.075cm);
\draw[thick, fill=black] (8,2) circle (0.075cm);
\draw[thick, fill=black] (8,1) circle (0.075cm);
\draw[thick, fill=black] (8,0) circle (0.075cm);
\draw[thick, fill=black] (8,-1) circle (0.075cm);
\draw[thick, fill=black] (8,-2) circle (0.075cm);
\draw[thick, blue, fill=black] (8,-3) circle (0.075cm);
\draw[thick, fill=black] (8,-4) circle (0.075cm);
\draw[thick,<->] (8.4,-4) -- (8.4,-3.05);
\node at (8.6,-3.5) {\small$\tau$};
\node at (7.6,-3) {\small $T_0$};
\node at (7.5,1) {\small $T_1$};
\draw[thick,red,<->] (8.4,-2.95) -- (8.4,1.45);
\node[red] at (8.7,-1) {\small$t$};
\draw[red] (8,-3) -- (8.6,-3);
\draw[red] (8,1.5) -- (8.6,1.5);
\node[blue!50!white] at (7.3,2.2) {\small $\Delta T$};
\end{tikzpicture}
}}
\caption{Coarse Graining of the time and space-variables.}
\label{Fig:CoarseGraining}
\end{center}
\end{figure}

Instead of the probability of finding the walker at the fixed lattice position $\vec{X}_1$ at time $T_1$ (assuming that they started at $\vec{X}_0=\vec{X}_1-\vec{X}$ at time $T_0=T_1-T$), we consider the probability to find the walker during a time interval $\Delta T$  in a region $\Delta X$ (with $\vec{X}_1\in \Delta X$). As schematically shown in Figure~\ref{Fig:CoarseGraining}, we consider $\Delta T$ to be labelled by a continuous time variable $t\in\mathbb{R}_+$ and $\Delta X$ by (Euclidean) coordinates\footnote{For convenience, we choose a parametrisation, such that $\vec{x}=0$ parametrises a $\Delta X$ that includes $\vec{X}_0$.} $\vec{x}=(x_1,\ldots, x_d)\in\mathbb{R}^d$ such that the length of the time interval $|\Delta T|$ is fixed and independent of $t$ and the volume of $\Delta X$ is independent of $\vec{x}$. Notice, while in general\footnote{Since here we are ultimately interested in the limit $a\to 0$ we can allow $\Delta X$ to be labelled by a coordinate that does not necessarily represent a lattice site itself.} $\vec{x}\notin \Gamma$ it can still be expanded in a series expansion in the basis vectors $\vec{E}_i$ (or equivalently $\vec{e}_i$)
\begin{align}
&\vec{x}=\sum_{i=1}^d x_i\,\vec{e}_i=a\,\sum_{i=1}^d\,X_i\,\vec{e}_i=\sum_{i=1}^d\,X_i\, E_i\,,&&\text{with} && &&x_i=a\,X_i\in\mathbb{R}\,&&\forall i=1,\ldots,d\,.
\end{align}
Furthermore, in the following we shall assume that $\Delta T$ and $\Delta X$ have the following properties

\begin{itemize}
\item[{\emph{(i)}}] the length $|\Delta T|$ of the interval $\Delta T$ is large compared to $\tau$ and the volume of $\Delta X$ (measured in $\mathbb{R}^d$) is large compared to the volume of the unit cell of $\Gamma$
\begin{align}
&|\Delta T|\gg \tau\,,&&\text{and} &&|\Delta X|=\text{vol}_{\mathbb{R}^d}(\Delta X)\gg a^d\,.\label{DefDeltaDiffusion}
\end{align} 
\item[{\emph{(ii)}}] the probability to find the walker does not vary too much over $\Delta T$ and $\Delta X$ (which is satisfied in general if $\frac{T}{\tau}=\frac{T_1-T_0}{\tau}\gg 1$)
\end{itemize} 
Both assumptions imply that we coarse grain the system at a time at which the walker has performed sufficiently many steps such that the probabilities $P(\vec{X},T)$ are slowly varying functions on $\Gamma$ which can be considered constant over the scales given by the volumes $|\Delta T|$ and $|\Delta X|$. These are in fact exactly the requirements for the system to be in a state that was called \emph{intermediate asymptotics} in \cite{Barenblatt} (see Apendix~\ref{App:IntermediateAsymptotics}), \emph{i.e.} the system being in a state which no longer (strongly) depends on the precise details of the initial conditions but has not yet settled into a static configuration. 
Furthermore, the assumptions \emph{(i)} and \emph{(ii)} allow us to introduce the following probability density
\begin{align}
p_{a,\tau}(\vec{x},t)|\Delta X|=\sum_{\vec{X}\in \Delta X}P(\vec{X},T)\sim\frac{|\Delta X|}{a^d}P(\vec{X},T)\,,\label{ProbDensityFinite}
\end{align}
which is implicitly a function of $a$ and $\tau$.

\subsubsection{Continuum Limit}
We are interested in the limit where both the lattice spacing $a$ and the time interval $\tau$ tend to zero, thus leading to a continuum theory. We are thus lead to consider
\begin{align}
p(\vec{x},t):=\lim_{a,\tau\to 0}\frac{1}{(2\pi)^d}\,\int_{\left[-\frac{\pi}{a},\frac{\pi}{a}\right]^d} d^d k\, e^{i\sum_{i=1}^dk_ix_i}\,\left(\frac{1}{d}\sum_{j=1}^d\cos(a\,k_j)\right)^{\frac{t}{\tau}}\,.\label{ProbabilityDensity}
\end{align}
This expression, however, strongly depends on the way the limit $a,\tau\to 0$ is taken. To understand this, we consider  (for $\alpha\in\mathbb{R}$ and $\beta\in\mathbb{R}_+$)
\begin{align}
\lim_{u\to 0^+}\left(\cos(u \alpha)\right)^{\frac{\beta}{u^p}}=\lim_{u\to 0^+}\left(1-\frac{u^2\alpha^2}{2}+\mathcal{O}(u^4)\right)^{\frac{\beta}{u^p}}=\left\{\begin{array}{lcl}1 & \text{if} & 0<p<2\,, \\ e^{-\frac{\beta\alpha^2}{2}} & \text{if} & p=2\,, \\ 0 & \text{if} & p>2\,.\end{array}\right. \label{LimitCos}
\end{align}
Therefore, we define the limit in (\ref{ProbabilityDensity}) as
\begin{align}
&\tau\rightarrow 0\,,&&\text{and}&&a\rightarrow 0\,,&&\text{such that} &&\frac{\tau}{a^2}\rightarrow \frac{1}{2d\kappa}\hspace{0.3cm}(\text{finite})\,,\label{FinLim}
\end{align}
where $\kappa\in\mathbb{R}\setminus\{0\}$ is an arbitrary constant, such that
\begin{align}
\lim_{a,\tau\to 0}\left(\frac{1}{d}\sum_{j=1}^d\cos(a\,k_j)\right)^{\frac{t}{\tau}}\longrightarrow\,\lim_{u\to 0}\left(\frac{1}{d}\sum_{j=1}^d\cos\left(u\,k_j\right)\right)^{\frac{2d \kappa t}{u^2}}
=\text{exp}\left(-\kappa t\sum_{j=1}^dk_j^2\right)\,.\label{LimitCosSummed}
\end{align}
We therefore find for the probability density (\ref{ProbabilityDensity}) the following Gaussian integral
\begin{align}
&p(\vec{x},t)=\frac{1}{(2\pi)^d}\int_{\mathbb{R}^d} d^d k \, e^{-\sum_{j=1}^d(\kappa t k_j^2-ik_j x_j)}=\frac{e^{-\frac{|\vec{x}|^2}{4\kappa t}}}{\left(4\pi \kappa t\right)^{d/2}}\,,&&\forall t>0\,.\label{CausalGreensFunction}
\end{align}
We remark that this probability density is normalised such that $\forall t>0$
\begin{align}
\int_{\mathbb{R}^d}d^dx\, p(\vec{x},t)=1\,,
\end{align}
and it furthermore satisfies the following Markov property 
\begin{align}
\int_{\mathbb{R}^d} d^d x\,p(\vec{x}_2-\vec{x},t_2-t_1)\,p(\vec{x}-\vec{x}_0,t_1-t_0)=\,p(\vec{x}_2-\vec{x}_0,t_2-t_0)\,,&&\forall t_2>t_1>t_0\,.
\end{align}
To get a better physical understanding of the result (\ref{CausalGreensFunction}) it is instructive to calculate $\langle|\vec{x}|\rangle$ (\emph{i.e.} the expectation value of the distance from the origin that the walker has moved) as a function of time
\begin{align}
\langle |\vec{x}|\rangle(t)&=\int_{\mathbb{R}^d}d^dx'\,|\vec{x}\,'|\, p(\vec{x}\,',t)=2\frac{\Gamma\left(\frac{1+d}{2}\right)}{\Gamma\left(\frac{d}{2}\right)}\,\sqrt{\kappa t}\,.\label{ExpValueRandomWalk}
\end{align}

\noindent
The fact that this expectation value behaves as $t^{1/2}$ is the reason why only $p=2$ in (\ref{LimitCos}) leads to a non-trivial limit (\ref{FinLim}) for the probability density. We can see from Figure~\ref{Fig:RandomWalks} that this behaviour is already encoded in (\ref{PFourierInt}) and is not a consequence of the limit (\ref{FinLim}): in the case $d=1$, the left panel shows the probabilities (\ref{PFourierInt}) for different times $T$ which are fitted with Gaussian functions, whose width in turn is fitted by $\Delta P=\sqrt{\frac{T}{\tau}}$ in the right part of the Figure.

\subsubsection{Heat Equation}\label{Sect:HeatEq}
In the previous Subsubsection we have taken the continuum limit at the level of the probability density (\ref{ProbabilityDensity}). We can equally discuss this limit at the level of the recursion relation (\ref{RecursionRelation}) which implies for the probability density 
\begin{align}
p_{a,\tau}(\vec{x},t+\tau)-p_{a,\tau}(\vec{x},t)=\frac{1}{2d}\,\sum_{i=1}^d \left[p_{a,\tau}(\vec{x}+ a\,\vec{e}_i,t) + p_{a,\tau}(\vec{x} -a\,\vec{e}_i,t) - 2p_{a,\tau}(\vec{x},t) \right]\,.\label{DiffEqProbDens}
\end{align}
Expanding the left hand side as a series in $\tau$ and the right hand side as a series in $a$ we find
\begin{align}
\tau\,\frac{\partial p_{a,\tau}}{\partial t}(\vec{x},t)+\mathcal{O}(\tau^2)&=\frac{a^2}{2d}\sum_{i,j,k=1}^d(\vec{e}_i)_j\,(\vec{e}_i)_k \frac{\partial^2 p_{a,\tau}}{\partial x_j \partial x_k}(\vec{x},t)+\mathcal{O}(a^3)\,.\label{ExpansionHeat}
\end{align}
Since $\{\vec{e}_1,\ldots,\vec{e}_d\}$ is a complete (orthonormalised) basis of $\mathbb{R}^d$, we have $\sum_{i=1}^d(\vec{e}_i)_j\,(\vec{e}_i)_k=\delta_{jk}$. Furthermore, we divide (\ref{ExpansionHeat}) by $\tau$ and implement the limit (\ref{FinLim}) by first identifying $\tau=\frac{a^2}{2d\kappa}$ and by defining
\begin{align}
p_a(\vec{x},t):=p_{a,\tau=\frac{a^2}{2d\kappa}}(\vec{x},t)=\frac{1}{(2\pi)^d}\,\prod_{i=1}^d\int_{\left[-\frac{\pi}{a},\frac{\pi}{a}\right]} d k_i\, e^{ik_ix_i}\,\left(\frac{1}{d}\sum_{j=1}^d\cos(a\,k_j)\right)^{\frac{2d\kappa t}{a^2}}\,.\label{paExpansion}
\end{align}
The limit $a\to 0$ then becomes
\begin{align}
\lim_{a\to 0}\,\frac{\partial p_{a}}{\partial t}(\vec{x},t)=\lim_{a\to 0}\left[\kappa\,\sum_{j=1}^d\,\frac{\partial^2 p_{a}}{\partial x_j^2}(\vec{x},t)+\mathcal{O}(a)\right]\,.\label{LimLaplace}
\end{align}
Upon introducing the Laplace operator $\Delta_{\vec{x}}=\sum_{i=1}^d\frac{\partial^2}{\partial x_i^2}$ on $\mathbb{R}^d$ and using (\ref{ProbabilityDensity}), the limit in (\ref{LimLaplace}) becomes (for $t>0$)
\begin{align}
(\partial_t-\kappa\, \Delta_{\vec{x}})\,p(\vec{x},t)=0\,,\label{DefHeatEq}
\end{align}
which is the heat equation with (thermal) diffusivity $\kappa$. This result is indeed consistent with the fact that (\ref{CausalGreensFunction}) is the causal Green's function of the differential operator $(\partial_t-\kappa\,\Delta_{\vec{x}})$ on $\mathbb{R}^d\times \mathbb{R}$ and with the boundary condition
\begin{align}
&\lim_{|\vec{x}|\rightarrow\infty}\,p(\vec{x},t)=0\,,&&\forall t>0\,.
\end{align}

\subsection{Perturbation around the Continuum Limit}
While we have seen in the previous Subsection that the limit (\ref{FinLim}) leads to a well-defined and non-trivial probability density $p(\vec{x},t)$, in this Subsection we are interested in perturbations around this limit. To this end, we treat $a$ (and thus also $\tau=\frac{a^2}{2d\kappa}$) as a small parameter and study the leading perturbations of (\ref{CausalGreensFunction}) and (\ref{DefHeatEq}).

\subsubsection{Perturbation of the Probability Density}
\begin{figure}[htbp]
\begin{center}
\includegraphics[width=7.5cm]{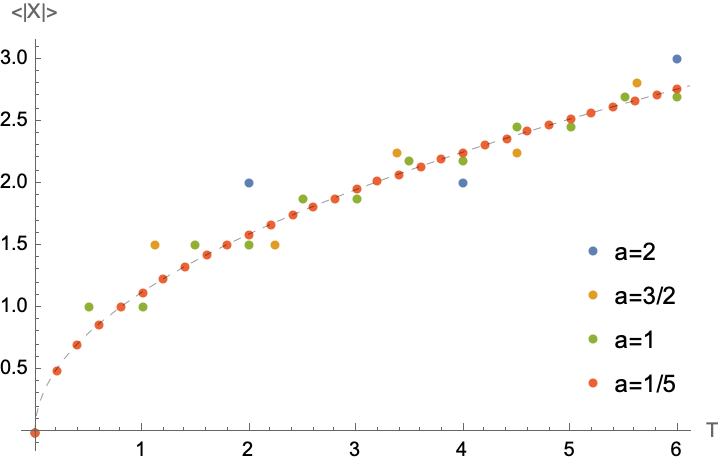}\hspace{1cm}\includegraphics[width=7.5cm]{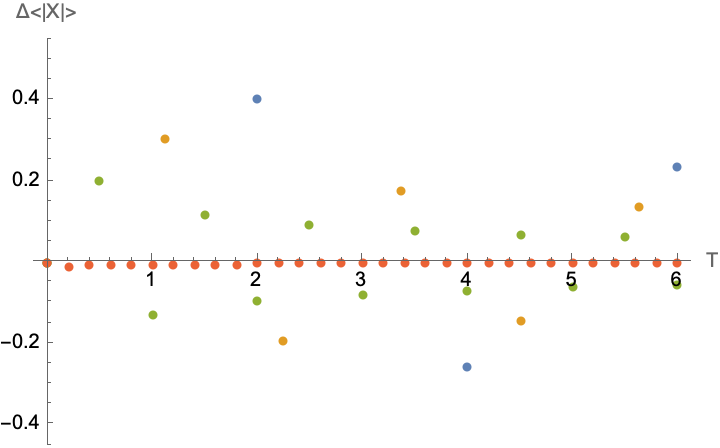}
\end{center}
\caption{Expectation value (\ref{DefQuantisedExpectation}) to leading order in $a$: left panel $\langle |\vec{X}|\rangle$ for $d=1$, $\kappa=1$ and different values of $a$. The thin dashed line represents the limit $a\to 0$ in (\ref{ExpValueRandomWalk}). Right panel: the difference $\langle |\vec{X}|\rangle-\langle |\vec{x}|\rangle$ representing the correction to the limit (\ref{ExpValueRandomWalk}). The values for $a$ are the same as in the left panel.}
\label{Fig:Perturbations}
\end{figure}

To obtain a first intuitive picture for the probabilities for small $a$, we return to the stochastic model described by the probability (\ref{PFourierInt}) and consider the expectation value
\begin{align}
\langle |\vec{X}|\rangle(T)=\sum_{\vec{X}\,'\in\Gamma}|\vec{X}\,'|\,P(\vec{X}\,',T)\big|_{\tau=\frac{a^2}{2d\kappa}}\,,\label{DefQuantisedExpectation}
\end{align}
For $d=1$, this function of $T$ (quantised in units of $\tau=\frac{a^2}{2d\kappa}$) is shown for different values of $a$ in Figure~\ref{Fig:Perturbations}. This plot suggests that the corrections to the expectation value (\ref{ExpValueRandomWalk}) are small and decrease as $T$ grows larger.

\begin{figure}[htbp]
\begin{center}
\includegraphics[width=7.5cm]{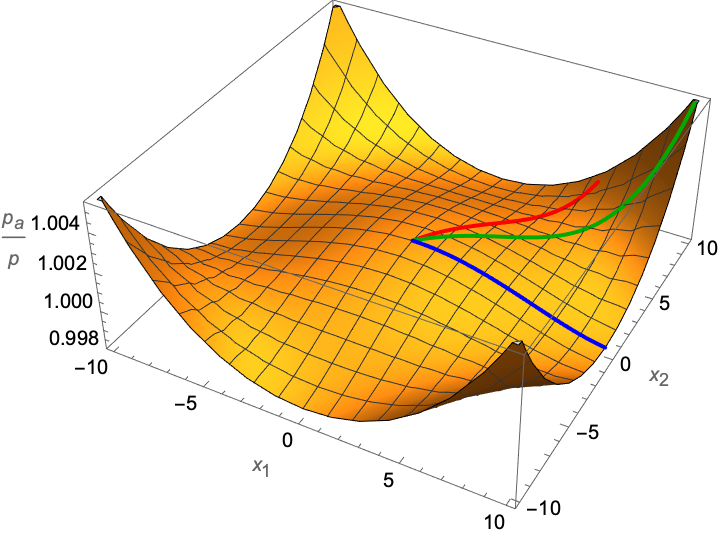}\hspace{1cm}\includegraphics[width=7.5cm]{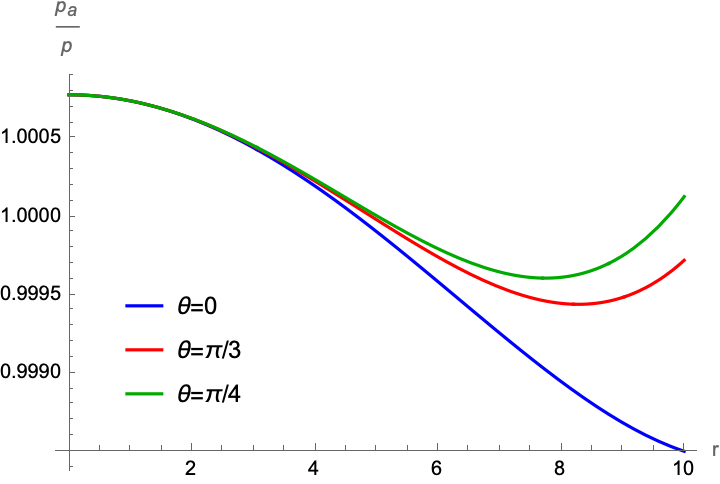}
\end{center}
\caption{Loss of spherical symmetry of the probability density (\ref{SubleadingDiffusion}) for $d=2$: the left panel shows $p_a/p$ as a function of $\vec{x}=(x_1,x_2)$ for $\kappa t=10$ and $a=1/2$. The coloured lines highlight the curves parametrised by $(r \cos\theta,r\sin\theta)$ for $\theta\in\{0,\pi/3,\pi/4\}$, which are re-plotted in the right panel for better comparison.}
\label{Fig:LossRotational}
\end{figure}

To obtain an analytic expression, we consider the deterministic description and (based on the definition (\ref{ProbDensityFinite})) study the family of probability densities (\ref{paExpansion})
for finite (but small) values of $a$. The leading correction to $p_a$ is calculated in Appendix~\ref{App:ErfExpansion} (see (\ref{PerturbGreen})), and can be written in the form
\begin{align}
p_a(\vec{x},t)&=p(\vec{x},t)\,\left[1-\frac{a^2}{192d(\kappa t)^3}\left(3|\vec{x}|^4-d\sum_{j=1}^d x_j^4-24 \kappa t |\vec{x}|^2+24 d \kappa^2 t^2\right)\right]+\mathcal{O}(a^4)\,.\label{SubleadingDiffusion}
\end{align}
While the limit (\ref{CausalGreensFunction}) is spherically symmetric, the leading correction breaks this symmetry. Indeed, in Figure~\ref{Fig:LossRotational} the subleading contribution is plotted for $d=2$ showing a breaking of the spherical symmetry. Finally, the subleading corrections to the expectation value (\ref{ExpValueRandomWalk}) are given by
\begin{align}
&\langle |\vec{x}|\rangle=2\,\frac{\Gamma\left(\frac{1+d}{2}\right)}{\Gamma\left(\frac{d}{2}\right)}\,\sqrt{\kappa t}\left[1+\frac{a^2}{8d(d+2)\kappa t}+\mathcal{O}(a^4)\right]\,,&&\forall t>\frac{a^2}{2d\kappa}\,.
\end{align}
In general, perturbations of the expectation value (\ref{ExpValueRandomWalk}) of the continuum theory are suppressed by powers of $\frac{a^2}{\kappa t}$. This means, even for finite values of $a$, the continuum theory provides a good approximation for times $t\gg \frac{a^2}{\kappa}$, in which case (\ref{DefHeatEq}) can be seen as an effective description.

\subsubsection{Perturbation of the Heat Equation}
Before closing this Section, we remark that we can compute sub-leading contributions not only directly to the probability densities as in (\ref{SubleadingDiffusion}), but also to the differential equation that describes them (\emph{i.e.} (\ref{DefHeatEq})). To this end, we formally expand the probability density as an even power series in $a$
\begin{align}
p_a(\vec{x},t)=p(\vec{x},t)+a^2\,p^{(2)}(\vec{x},t)+\mathcal{O}(a^4)\,.
\end{align}
Inserting this expansion into the difference equation (\ref{DiffEqProbDens}) (for $\tau=\frac{a^2}{2d\kappa}$) we can expand it order by order in $a$ to find the following equations
\begin{align}
&\text{order }\mathcal{O}(a^2):&&\frac{1}{2d\kappa}\left(\partial_t\, p(x,t)-\kappa\,\Delta\,p(x,t)\right)=0\,,\label{EqExpanddiff0}\\
&\text{order }\mathcal{O}(a^4):&&\frac{1}{2d\kappa}\left(\partial_t\, p^{(2)}(x,t)-\kappa\,\Delta\,p^{(2)}(x,t)\right)=\frac{1}{4!d}\sum_{i=1}^d\frac{\partial^4 p(\vec{x},t)}{\partial x_i^4}-\frac{1}{8d^2\kappa^2}\,\partial_t^2 p(\vec{x},t)\,.\label{EqExpanddiff2}
\end{align}
Here the equation (\ref{EqExpanddiff0}) at order $\mathcal{O}(a^2)$ is the result obtained in (\ref{ExpansionHeat}) and (\ref{DefHeatEq}) namely the heat equation. The solution of this equation $p(\vec{x},t)$ acts as inhomogeneity in the heat equation (\ref{EqExpanddiff2}) for the correction $p^{(2)}$. While both (\ref{EqExpanddiff0}) and (\ref{EqExpanddiff2}) a priori allow for more solutions, we have verified that for $p(x,t)$ as in (\ref{CausalGreensFunction}), the subleading contribution in (\ref{SubleadingDiffusion}) is indeed a solution of (\ref{EqExpanddiff2}).
\section{Epidemiology: Simple Epidemic- and the SI-Model}\label{Sect:SimpleEpidemicSI}
After discussing diffusion processes in the previous Section, we now consider a system that describes in a simplistic fashion the spread of a disease within a population. We shall not allow for the possibility of recovery. We first discuss a purely stochastic approach, which can be approximated by a deterministic model  similar to the one considered in \cite{Grassberger1983}.

\subsection{Stochastic Approach}\label{Sect:StochasticExamples}
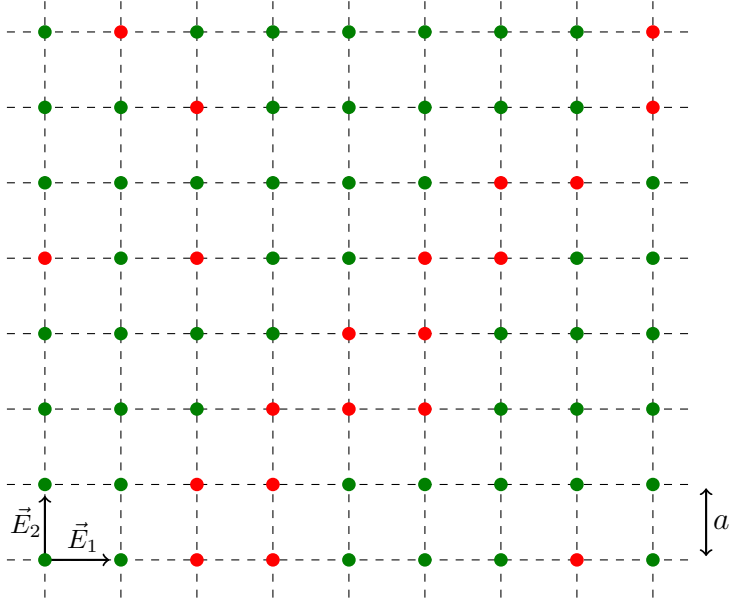
\begin{figure}[htbp]
\begin{center}
\scalebox{1}{\parbox{9.8cm}{\begin{tikzpicture}
%
\draw[dashed] (-4.5,3) -- (4.5,3);
\draw[dashed] (-4.5,2) -- (4.5,2);
\draw[dashed] (-4.5,1) -- (4.5,1);
\draw[dashed] (-4.5,0) -- (4.5,0);
\draw[dashed] (-4.5,-1) -- (4.5,-1);
\draw[dashed] (-4.5,-2) -- (4.5,-2);
\draw[dashed] (-4.5,-3) -- (4.5,-3);
\draw[dashed] (-4.5,-4) -- (4.5,-4);
\draw[dashed] (4,-4.5) -- (4,3.5);
\draw[dashed] (3,-4.5) -- (3,3.5);
\draw[dashed] (2,-4.5) -- (2,3.5);
\draw[dashed] (1,-4.5) -- (1,3.5);
\draw[dashed] (0,-4.5) -- (0,3.5);
\draw[dashed] (-1,-4.5) -- (-1,3.5);
\draw[dashed] (-2,-4.5) -- (-2,3.5);
\draw[dashed] (-3,-4.5) -- (-3,3.5);
\draw[dashed] (-4,-4.5) -- (-4,3.5);
\draw[thick,->] (-4,-4) -- (-3.15,-4);
\node at (-3.5,-3.7) {\footnotesize $\vec{E}_1$};
\draw[thick,->] (-4,-4) -- (-4,-3.15);
\node at (-4.25,-3.5) {\footnotesize $\vec{E}_2$};
\draw[thick, fill=green!50!black,green!50!black] (-4,3) circle (0.075cm);
\draw[thick, fill=red,red] (-3,3) circle (0.075cm);
\draw[thick, fill=green!50!black,green!50!black] (-2,3) circle (0.075cm);
\draw[thick, fill=green!50!black,green!50!black] (-1,3) circle (0.075cm);
\draw[thick, fill=green!50!black,green!50!black] (0,3) circle (0.075cm);
\draw[thick, fill=green!50!black,green!50!black] (1,3) circle (0.075cm);
\draw[thick, fill=green!50!black,green!50!black] (2,3) circle (0.075cm);
\draw[thick, fill=green!50!black,green!50!black] (3,3) circle (0.075cm);
\draw[thick, fill=red,red] (4,3) circle (0.075cm);
\draw[thick, fill=green!50!black,green!50!black] (-4,2) circle (0.075cm);
\draw[thick, fill=green!50!black,green!50!black] (-3,2) circle (0.075cm);
\draw[thick, fill=red,red] (-2,2) circle (0.075cm);
\draw[thick, fill=green!50!black,green!50!black] (-1,2) circle (0.075cm);
\draw[thick, fill=green!50!black,green!50!black] (0,2) circle (0.075cm);
\draw[thick, fill=green!50!black,green!50!black] (1,2) circle (0.075cm);
\draw[thick, fill=green!50!black,green!50!black] (2,2) circle (0.075cm);
\draw[thick, fill=green!50!black,green!50!black] (3,2) circle (0.075cm);
\draw[thick, fill=red,red] (4,2) circle (0.075cm);
\draw[thick, fill=green!50!black,green!50!black] (-4,1) circle (0.075cm);
\draw[thick, fill=green!50!black,green!50!black] (-3,1) circle (0.075cm);
\draw[thick, fill=green!50!black,green!50!black] (-2,1) circle (0.075cm);
\draw[thick, fill=green!50!black,green!50!black] (-1,1) circle (0.075cm);
\draw[thick, fill=green!50!black,green!50!black] (0,1) circle (0.075cm);
\draw[thick, fill=green!50!black,green!50!black] (1,1) circle (0.075cm);
\draw[thick, fill=red,red] (2,1) circle (0.075cm);
\draw[thick, fill=red,red] (3,1) circle (0.075cm);
\draw[thick, fill=green!50!black,green!50!black] (4,1) circle (0.075cm);
\draw[thick, fill=red,red] (-4,0) circle (0.075cm);
\draw[thick, fill=green!50!black,green!50!black] (-3,0) circle (0.075cm);
\draw[thick, fill=red,red] (-2,0) circle (0.075cm);
\draw[thick, fill=green!50!black,green!50!black] (-1,0) circle (0.075cm);
\draw[thick, fill=green!50!black,green!50!black] (0,0) circle (0.075cm);
\draw[thick, fill=red,red] (1,0) circle (0.075cm);
\draw[thick, fill=red,red] (2,0) circle (0.075cm);
\draw[thick, fill=green!50!black,green!50!black] (3,0) circle (0.075cm);
\draw[thick, fill=green!50!black,green!50!black] (4,0) circle (0.075cm);
\draw[thick, fill=green!50!black,green!50!black] (-4,-1) circle (0.075cm);
\draw[thick, fill=green!50!black,green!50!black] (-3,-1) circle (0.075cm);
\draw[thick, fill=green!50!black,green!50!black] (-2,-1) circle (0.075cm);
\draw[thick, fill=green!50!black,green!50!black] (-1,-1) circle (0.075cm);
\draw[thick, fill=red,red] (0,-1) circle (0.075cm);
\draw[thick, fill=red,red] (1,-1) circle (0.075cm);
\draw[thick, fill=green!50!black,green!50!black] (2,-1) circle (0.075cm);
\draw[thick, fill=green!50!black,green!50!black] (3,-1) circle (0.075cm);
\draw[thick, fill=green!50!black,green!50!black] (4,-1) circle (0.075cm);
\draw[thick, fill=green!50!black,green!50!black] (-4,-2) circle (0.075cm);
\draw[thick, fill=green!50!black,green!50!black] (-3,-2) circle (0.075cm);
\draw[thick, fill=green!50!black,green!50!black] (-2,-2) circle (0.075cm);
\draw[thick, fill=red,red] (-1,-2) circle (0.075cm);
\draw[thick, fill=red,red] (0,-2) circle (0.075cm);
\draw[thick, fill=red,red] (1,-2) circle (0.075cm);
\draw[thick, fill=green!50!black,green!50!black] (2,-2) circle (0.075cm);
\draw[thick, fill=green!50!black,green!50!black] (3,-2) circle (0.075cm);
\draw[thick, fill=green!50!black,green!50!black] (4,-2) circle (0.075cm);
\draw[thick, fill=green!50!black,green!50!black] (-4,-3) circle (0.075cm);
\draw[thick, fill=green!50!black,green!50!black] (-3,-3) circle (0.075cm);
\draw[thick, fill=red,red] (-2,-3) circle (0.075cm);
\draw[thick, fill=red,red] (-1,-3) circle (0.075cm);
\draw[thick, fill=green!50!black,green!50!black] (0,-3) circle (0.075cm);
\draw[thick, fill=green!50!black,green!50!black] (1,-3) circle (0.075cm);
\draw[thick, fill=green!50!black,green!50!black] (2,-3) circle (0.075cm);
\draw[thick, fill=green!50!black,green!50!black] (3,-3) circle (0.075cm);
\draw[thick, fill=green!50!black,green!50!black] (4,-3) circle (0.075cm);
\draw[thick, fill=green!50!black,green!50!black] (-4,-4) circle (0.075cm);
\draw[thick, fill=green!50!black,green!50!black] (-3,-4) circle (0.075cm);
\draw[thick, fill=red,red] (-2,-4) circle (0.075cm);
\draw[thick, fill=red,red] (-1,-4) circle (0.075cm);
\draw[thick, fill=green!50!black,green!50!black] (0,-4) circle (0.075cm);
\draw[thick, fill=green!50!black,green!50!black] (1,-4) circle (0.075cm);
\draw[thick, fill=green!50!black,green!50!black] (2,-4) circle (0.075cm);
\draw[thick, fill=red,red] (3,-4) circle (0.075cm);
\draw[thick, fill=green!50!black,green!50!black] (4,-4) circle (0.075cm);
\draw[<->,thick] (4.7,-3.95) -- (4.7,-3.05);
\node at (4.9,.-3.5) {$a$};
\end{tikzpicture}
}}
\caption{Configuration of infectious (red) and susceptible (green) individuals as sites on the lattice $\Gamma$ (for $d=2$) generated by the basis vectors $(\vec{E}_1,\vec{E}_2)$.}
\label{Fig:EpidemicPopulation}
\end{center}
\end{figure}

We consider a(n infinite) population whose inhabitants are represented as the lattice sites on the $d$-dimensional infinite, hypercubic lattice $\Gamma$ defined in eq.~(\ref{DefHyperLattice}). At any given time $T$, we separate the population into two compartments, namely each individual is either \emph{infectious} (\emph{i.e.} they carry a certain communicable disease) or \emph{susceptible} (see Figure~\ref{Fig:EpidemicPopulation} for a sample configuration in $d=2$).\footnote{As is customary in epidemiology, a susceptible individual is currently not infected with the disease, however, upon contact with an infectious individual may contract the disease and become infectious themselves (\emph{i.e.} capable of infecting other susceptible). At this stage, we do not consider the possibility of recovery, \emph{i.e.} an infectious individual will remain so forever. We plan to come back to this point in later work.} We assume that in fixed time-intervals $\tau$ infectious individuals can, with a certain probability, infect nearby susceptible individuals (such that they turn into infectious individuals themselves). To this end, we introduce the function
\begin{align}
&\parbox{3.7cm}{
$F:\hspace{0.5cm}\Gamma\longrightarrow [0,1/\tau]$
}
&&\text{with}&&0\leq \gamma':=\sum_{\vec{X}\in\Gamma}F(\vec{X})<\infty\,,\label{SIConditionFunctionPre}
\end{align}
which defines the rate at which an individual at $\vec{X}$ infects susceptible individuals at $\vec{X}'$.\footnote{For simplicity, we assume in this work that this rate is translational invariant, \emph{i.e.} it is only a function of the difference $\vec{X}-\vec{X}\,'$.} Furthermore, let $P(\vec{X},T)$ denote the probability for the individual at the lattice site $\vec{X}$ to be infectious at time $T\in\mathbb{R}$, then we have the following recursion relation

\begin{align}
P(\vec{X},T+\tau)=P(\vec{X},T)+\tau\,\left(1-P(\vec{X},T)\right)\,\sum_{\vec{X}\,'\in\Gamma}F(\vec{X}-\vec{X}')\,P(\vec{X}\,',T)\,.\label{StochasticEvolution}
\end{align}
Starting from an initial lattice configuration of susceptible and infectious individuals at some fixed $T_0$, this relation allows us to iteratively compute the probability for an individual at any $\vec{X}\in\Gamma$ to be infectious at a time $T_0+n\,\tau$ for any $n\in\mathbb{N}$. Notice, if $P(\vec{X},T)=1$ it will remain $1$ for all $T'\geq T$.

For later use and in order to gain some intuition for this model, we shall discuss some examples for the particular case $d=1$ numerically. In particular, we shall consider different initial conditions and choices of the function~$F$, that governs the way in which individuals infect one another:\footnote{For simplicity, and to simplify the notation in $d=1$, we shall denote the positions simply by $X$ (instead of~$\vec{X}$).}

\begin{wrapfigure}{r}{0.42\textwidth}
${}$\\[-1cm]
\begin{center}
\includegraphics[width=6.6cm]{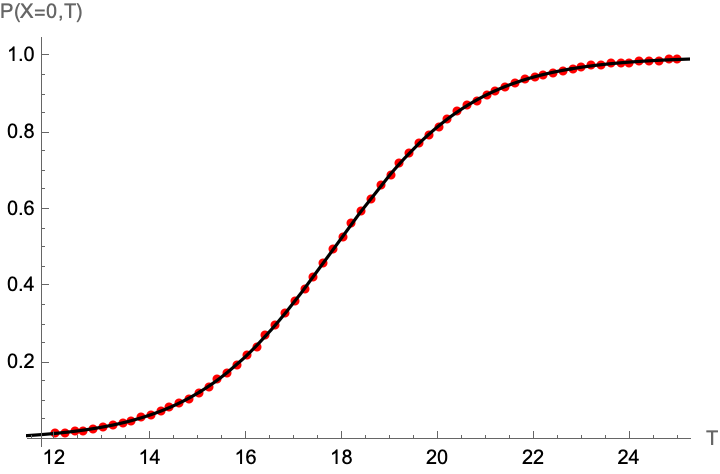}
\end{center}
\caption{Probabilities $P(X,T)$ in $d=1$ for the initial conditions~(\ref{DefHomogeneousInit}). The red dots show the solution (\ref{StochasticEvolution}) for $P_0=5\cdot 10^{-6}$, while the black line gives an interpolation by a logistic function.}
\label{Fig:HomogeneousDistr}
${}$\\[-1cm]
\end{wrapfigure}

${}$

\begin{itemize}
\item {\bf homogeneous distribution:} the simplest initial condition is to assume that at some time $T_0$ all individuals have the same probability to be infectious\footnote{Since we use $d=1$, we denote the spatial variable simply with $X\in a\mathbb{Z}$.}
\begin{align}
P(X,T=T_0)=P_0\hspace{1cm}\forall X\in\Gamma\cong \mathbb{Z}\,.\label{DefHomogeneousInit}
\end{align}
In this case, the time evolution shows a (trivial) self-similarity in the spatial distribution: the probabilities are the same at each lattice site, such that they become independent of $X$. This behaviour is still (approximately) observed if the initial probabilities $P_0$ are perturbed by a small $X$-dependent function. Schematically, the iterative solution of (\ref{StochasticEvolution}) is plotted in Figure~\ref{Fig:HomogeneousDistr} for a 1-dimensional configuration. While the plot was computed using a specific $F$ (the one in (\ref{ExpDecayNoSelf})), the result is independent of the details of this choice of $F$ but only depends on $\gamma'$ in (\ref{SIConditionFunctionPre}). Indeed, the solid black line in Figure~\ref{Fig:HomogeneousDistr} is an interpolation by a logistic function $P(0,T)=\frac{1}{1+e^{-\lambda(T-T_0)}}$ with $\lambda\sim\gamma'\sim0.6927$. As is showcased in Figure~\ref{Fig:HomogeneousDistr}, the time evolution in each point can be well approximated with this logistic growth.
\end{itemize}

\begin{itemize}
\item {\bf travelling waves:} we consider a starting configuration consisting of a number of infectious individuals placed (symmetrically) around the origin, \emph{i.e.} for $T_0=0$ we have
\begin{align}
P(X,T_0)=\left\{\begin{array}{lcl}1 & \text{if} & |X|\leq R_0\,,\\ 0 & \text{if} & |X|> R_0\end{array}\right.\label{StochasticInitialCond}
\end{align}

\begin{figure}[htbp]
\begin{center}
\includegraphics[width=7.5cm]{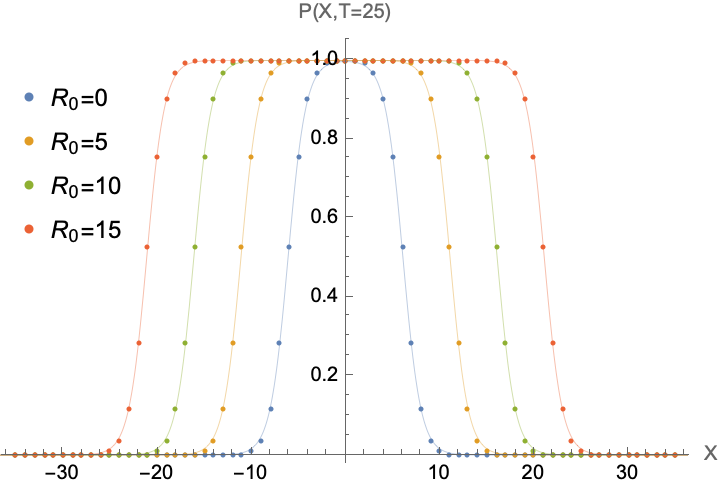}\hspace{1cm}\includegraphics[width=7.5cm]{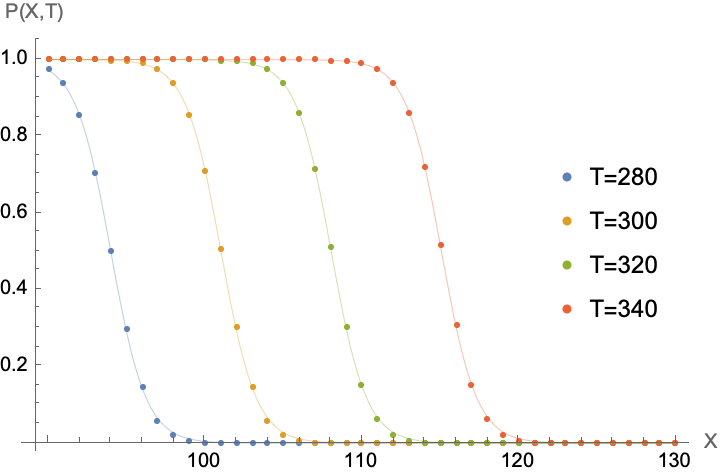}
\end{center}
\caption{Numerical evaluation of the probabilities following (\ref{StochasticEvolution}) with the initial conditions (\ref{StochasticInitialCond}) for $d=1$ and with $F(X,X')$ given in (\ref{ExampleConstantF}). The left panel shows the dependence of the evolution on the value of $R_0$ (for early times), while the right panel shows the evolution in the form of a traveling wave at different times for fixed $R_0=0$. Both plots use $\tau=1$ and $\gamma=0.15$.}
\label{Fig:WavesStochastic}
\end{figure}

\noindent
Figure~\ref{Fig:WavesStochastic} shows the numerical evaluation of the probability for an individual at $X$ to be infected at time $T$ for $d=1$ and for $F=F_{\text{adj}}$:
\begin{align}
F_{\text{adj}}(X)=\left\{\begin{array}{ll} \kappa & \text{if }X=\pm a\,, \\ 0 & \text{else}\end{array}\right.\label{ExampleConstantF}
\end{align}
with $\kappa=\in[0,\frac{1}{\tau}]$ a constant. This means, in this model only nearest neighbours can infect one another. Both plots in \ref{Fig:WavesStochastic} indicate that also in this case the  probability exhibits a self-similarity behaviour for large enough $T$, namely in the form of a travelling wave. Indeed, the solid lines in Figure~\ref{Fig:WavesStochastic} represent interpolations of the form
\begin{align}
P(X,T)=\frac{1}{1+e^{-\lambda(T)(X+X_0(T))}}-\frac{1}{1+e^{-\lambda(T)(X-X_0(T))}}\,.\label{StochasticApproximationD1Wave}
\end{align}
Since (due to the choice of initial conditions (\ref{StochasticInitialCond})), the time evolution is symmetric with respect to $X\to -X$, it is sufficient to focus on $X>0$, in which case, one can equivalently (for sufficiently large $T$) approximate the solution by the function
\begin{align}
&P(X,T)=1-\frac{1}{1+e^{-\lambda(T)(X-X_0(T))}}\,,&&\forall X>0\,.\label{ApproxStochastic2}
\end{align}
Here and in (\ref{StochasticApproximationD1Wave}), $\lambda,X_0:\mathbb{R}\to \mathbb{R}_+$ are functions of time, which are plotted in Figure~\ref{Fig:WavesStochasticParameters}. These suggest that for sufficiently large $T$ (\emph{i.e.} once the system has settled into a state of intermediate asymptotics), $\gamma$ approaches a constant (that only depends on $\lambda$), while $X_0$ tends to a linear function in time. This means that the travelling wave moves approximately with a constant velocity, which is in contrast to the behaviour of the expectation value of the position of a random walker found in (\ref{ExpValueRandomWalk}).

\begin{figure}[htbp]
\begin{center}
\includegraphics[width=7.5cm]{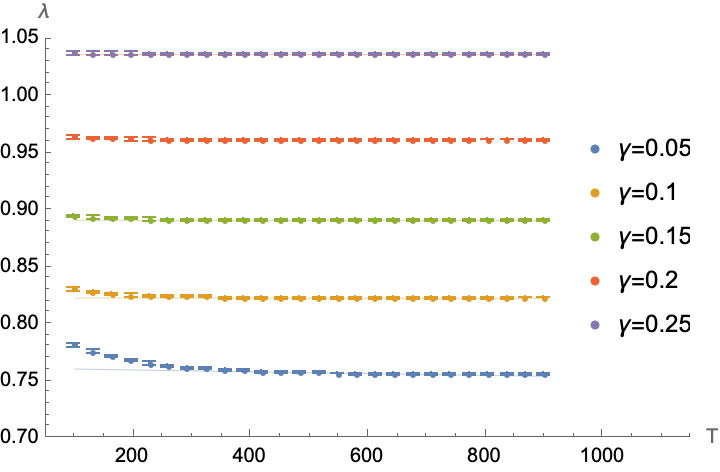}\hspace{1cm}\includegraphics[width=7.5cm]{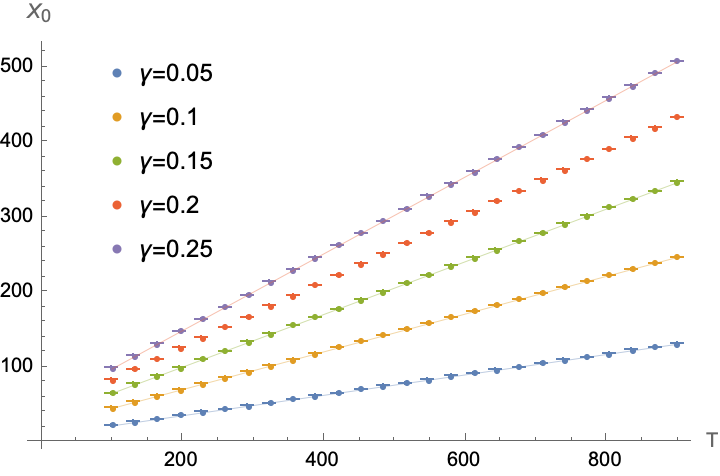}
\end{center}
\caption{Numerical plots of the functions $\lambda$ (left panel) and $X_0$ (right panel) appearing in the interpolating function (\ref{StochasticApproximationD1Wave}) as functions of time.}
\label{Fig:WavesStochasticParameters}
\end{figure}

\item {\bf periodic initial conditions:}
Another class of initial conditions are periodically fluctuating ones of the form
\begin{align}
P(X,T_0)=P_0\,(1+\cos(L X))\,,\label{StochasticInitialCondPeriod}
\end{align}
which are schamtically plotted in the left panel of Figure~\ref{Fig:PeriodicBoundary}. Such conditions could be used to mimic outbreak centers that are geographically spaced out with a characteristic length scale $L$.

\begin{figure}[htbp]
\begin{center}
\includegraphics[width=7.5cm]{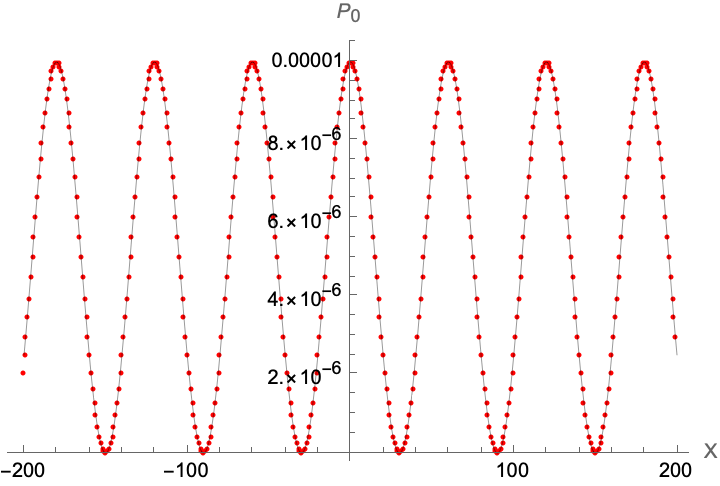}\hspace{1cm}\includegraphics[width=7.5cm]{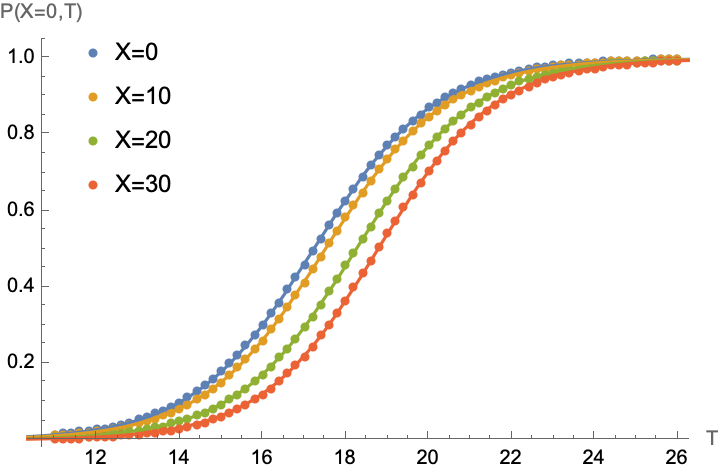}
\end{center}
\caption{Numerical evaluation of the probabilities following (\ref{StochasticEvolution}) with periodic initial conditions (\ref{StochasticInitialCondPeriod}) for $P_0=5\cdot 10^{-6}$ and $L=\frac{\pi}{30}$. The left panel shows the initial conditions, while the right panel shows the evolution of $P(X,T)$ as a function of $T$ for different fixed $X$ (for $F$ as in (\ref{ExpDecayNoSelf})). The solid lines give interpolations with logistic functions.}
\label{Fig:PeriodicBoundary}
\end{figure}

Here we  choose the  $F(X)$ function to be
\begin{align}
F(X)=\left\{\begin{array}{lcl}\frac{e^{-X^2/20}}{10} & \text{if} &X\neq 0\,, \\ 0 & \text{if} & X=0\,.\end{array}\right.\label{ExpDecayNoSelf}
\end{align}

The iterative solution of equation (\ref{StochasticEvolution}) is shown in the right panel of Figure~\ref{Fig:PeriodicBoundary} for different values of $X$. Each of these solutions can be very well approximated by a logistic function, however, with parameters that depend on $X$
\begin{align}
&P(X,T)\sim\frac{1}{1+e^{-\lambda(X)(T-T_0(X))}}\,\label{ApproxStochasticT}
\end{align}
Schematic plots of $\lambda(X)$ and $T_0(X)$ are shown in Figure~\ref{Fig:PeriodicBoundaryValuesLog}.

\begin{figure}[htbp]
\begin{center}
\includegraphics[width=7.5cm]{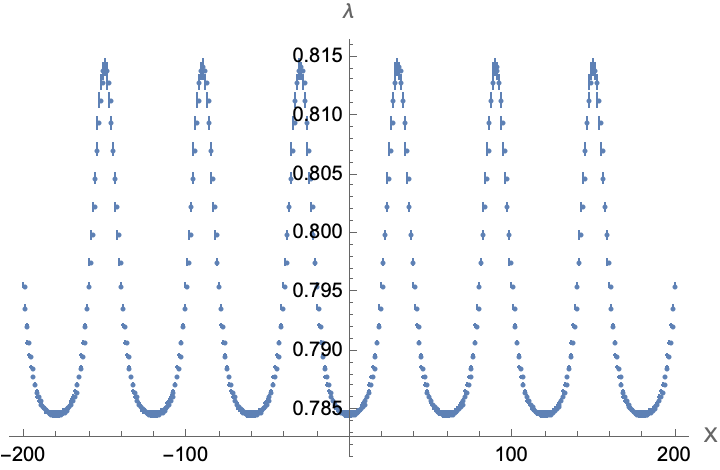}\hspace{1cm}\includegraphics[width=7.5cm]{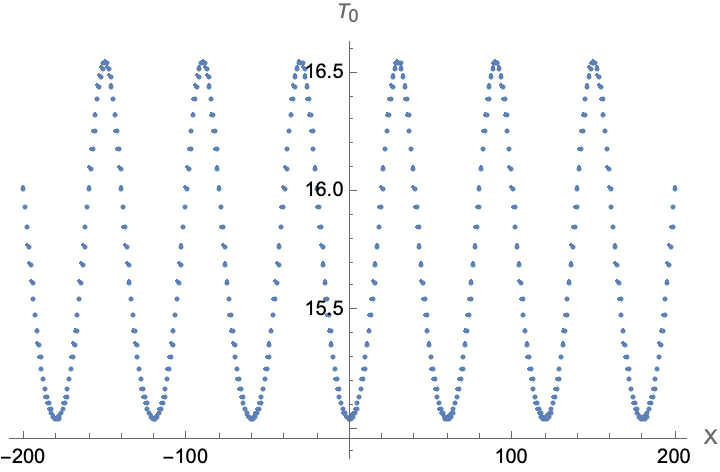}
\end{center}
\caption{Parameters $\lambda$ (left panel) and $T_0$ (right panel) of the approximations in (\ref{ApproxStochasticT}) for the probabilities computed from (\ref{StochasticEvolution}) with periodic boundary conditions (\ref{StochasticInitialCondPeriod}). Both plots use the same remaining parameters as in Figure~\ref{Fig:PeriodicBoundary}.}
\label{Fig:PeriodicBoundaryValuesLog}
\end{figure}

\end{itemize}

\subsection{Simple Epidemic Model}
Rather than working with the stochastic model (\ref{StochasticEvolution}), we shall ultimately work with a deterministic 'simple epidemic model' introduced in \cite{Mollison1977} (see also \cite{Kendall}). To this end, we shall approximate the former by replacing the time- and spatial variables by continuous ones.
\subsubsection{Deterministic Approximation: Continuous Time}
To motivate the relation to the stochastic approach discussed above, we first write the difference equation (\ref{StochasticEvolution}) in the following form
\begin{align}
\frac{P(\vec{X},T+\tau)-P(\vec{X},T)}{\tau}=\left(1-P(\vec{X},T)\right)\,\sum_{\vec{X}\,'\in\Gamma}F(\vec{X}-\vec{X}')\,P(\vec{X}\,',T)
\end{align}
and take the limit $\tau\to 0$
\begin{align}
&\frac{\partial P(\vec{X},t)}{\partial t}=\left(1-P(\vec{X},t)\right)\,\sum_{\vec{X}\,'\in\Gamma}F(\vec{X}-\vec{X}')\,P(\vec{X}\,',t)\,,&&\forall \vec{X}\in\Gamma\text{ and }t\in\mathbb{R}\,.\label{PropEquation}
\end{align}
Here we have furthermore replaced $T\to t\in\mathbb{R}$ and we consider the latter as a continuous variable (that is not quantised in units of $\tau$). This equation, however, can also be interpreted to come from a compartmental SI-model defined on the lattice $\Gamma$, similar to \cite{Grassberger1983}: to see this, we define
\begin{align}
&\mathfrak{S}:\hspace{0.5cm}\Gamma\times \mathbb{R}_+ \longrightarrow [0,1]\,,&& \mathfrak{I}:\hspace{0.5cm} \Gamma\times \mathbb{R}_+\longrightarrow [0,1]\,,\nonumber\\
&\hspace{1.35cm}(\vec{X},t)\longmapsto \mathfrak{S}(\vec{X},t)\,,&&\hspace{1.35cm}(\vec{X},t)\longmapsto \mathfrak{I}(\vec{X},t)\,,\label{DiscreteSIsys}
\end{align}
where $\mathfrak{S}$ and $\mathfrak{I}$ are interpreted as the (normalised) number of susceptible and infectious individuals respectively at the lattice site $\vec{X}$ and at the time $t$ (which is considered to be continuous).\footnote{For simplicity, we place the start of the epidemic at $t=0$.} In order to model the contact between susceptible and infectious individuals (and therefore also the infection dynamics), we define a function $F:\,\Gamma\longrightarrow \mathbb{R}_+$ (similar to (\ref{SIConditionFunctionPre})) with the property
\begin{align}
&\sum_{\vec{X}\in\Gamma}F(\vec{X})=\gamma'\,,&&\text{with} &&0<\gamma'<\infty\,,\label{SIConditionFunction}
\end{align}
and model the spread of the disease by the following differential equations in time
\begin{align}
\frac{\partial \mathfrak{S}}{\partial t}(\vec{X},t)&=-\mathfrak{S}(\vec{X},t)\,\sum_{\vec{X}\,'\in\Gamma}\,F(\vec{X}-\vec{X}\,')\,\mathfrak{I}(\vec{X}\,',t)\,,\nonumber\\
\frac{\partial \mathfrak{I}}{\partial t}(\vec{X},t)&=\mathfrak{S}(\vec{X},t)\,\sum_{\vec{X}\,'\in\Gamma}\,F(\vec{X}-\vec{X}\,')\,\mathfrak{I}(\vec{X}\,',t)\,,\label{EqDiffSimpleEpidemicDouble}
\end{align}
together with the initial conditions
\begin{align}
&\mathfrak{S}(\vec{X},0)=\mathfrak{S}_{\text{init}}(\vec{X})\,,&&\mathfrak{I}(\vec{X},0)=\mathfrak{I}_{\text{init}}(\vec{X})\,,&&\forall \vec{X}\in \Gamma\,,
\end{align}
where $\mathfrak{S}_{\text{init}}\,,\mathfrak{I}_{\text{init}}:\,\Gamma\longrightarrow [0,1]$ are functions that satisfy $\mathfrak{S}_{\text{init}}(\vec{X})+\mathfrak{I}_{\text{init}}(\vec{X})=1$ $\forall \vec{X}\in\Gamma$. Since (\ref{EqDiffSimpleEpidemic}) implies $\frac{d}{dt}\left(\mathfrak{S}+\mathfrak{I}\right)(\vec{X},t)=0$ we have
\begin{align}
&(\mathfrak{S}+\mathfrak{I})(\vec{X},t)=1\,,&&\forall\,\vec{X}\in\Gamma\,,\hspace{0.2cm} \forall t\in\mathbb{R}\,.\label{ConditionSum}
\end{align}
This allows us to write the coupled system of differential equations (\ref{EqDiffSimpleEpidemicDouble}) in the form of a single non-linear differential equation
\begin{align}
&\frac{\partial \mathfrak{I}}{\partial t}(\vec{X},t)=\left(1-\mathfrak{I}(\vec{X},t)\right)\,\sum_{\vec{X}\,'\in\Gamma}\,F(\vec{X}-\vec{X}\,')\,\mathfrak{I}(\vec{X}\,',t)\,.\label{EqDiffSimpleEpidemic}
\end{align}
This is the same equation as (\ref{PropEquation}), upon identifying directly the probabilities $P(\vec{X},t)$ for a lattice site $\vec{X}$ to be infected with the number of infectious individuals $\mathfrak{I}$ at $\vec{X}$. Identifying (\ref{EqDiffSimpleEpidemic}) with (\ref{PropEquation}) therefore marks the transition from a stochastic approach (characterised by probabilities) to a deterministic approach (with the number of infectious individuals as the observable).

The model (\ref{EqDiffSimpleEpidemic}) is motivated by the mass action law, which underlies many compartmental models, notably the SIR model \cite{Kermack:1927}: the rate at which susceptible individuals (at $\vec{X}$) are turned into infectious ones depends on the number of susceptible individuals at $\vec{X}$ and the number of infectious ones they can get into contact with. In (\ref{EqDiffSimpleEpidemic}) we are allowing for the possibilities that infectious from anywhere on $\Gamma$ may infect a susceptible at $\vec{X}$, however, the rate at which this is happening is weighted by the function $F$. Since the latter is bounded from below and $\mathfrak{I}(\vec{X},t)\in[0,1]$ $\forall t\in\mathbb{R}$, the condition (\ref{SIConditionFunction}) is a necessary condition for the equations (\ref{EqDiffSimpleEpidemic}) to be well-defined since
\begin{align}
0\leq \sum_{\vec{X}\,'\in\Gamma}\,F(\vec{X}-\vec{X}\,')\,\mathfrak{I}(\vec{X}\,',t)\leq \sum_{\vec{X}\,'\in\Gamma}\,F(\vec{X}-\vec{X}\,')=\gamma'<\infty\,.
\end{align}

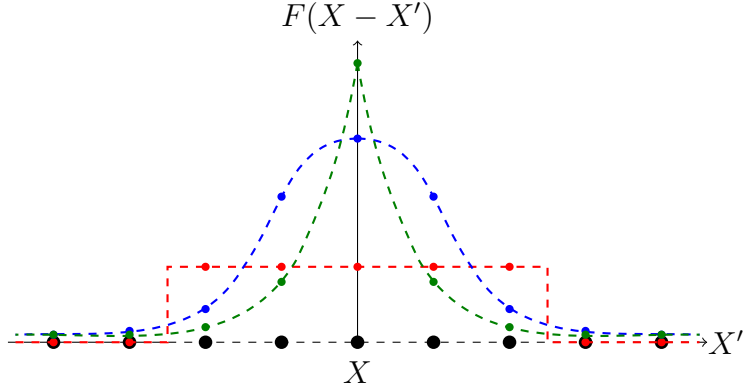
\begin{figure}
\begin{center}
\scalebox{1}{\parbox{9.7cm}{\begin{tikzpicture}
\draw[thick, fill=black] (-4,0) circle (0.075cm);
\draw[thick, fill=black] (-3,0) circle (0.075cm);
\draw[thick, fill=black] (-2,0) circle (0.075cm);
\draw[thick, fill=black] (-1,0) circle (0.075cm);
\draw[thick, fill=black] (0,0) circle (0.075cm);
\draw[thick, fill=black] (1,0) circle (0.075cm);
\draw[thick, fill=black] (2,0) circle (0.075cm);
\draw[thick, fill=black] (3,0) circle (0.075cm);
\draw[thick, fill=black] (4,0) circle (0.075cm);
\draw[dashed,->] (-4.6,0) -- (4.6,0);
\node at (0,4.3) {$F(X-X')$};
\node at (4.85,0) {$X'$};
\node at (0,-0.4) {$X$};
\draw[->] (0,0) -- (0,4);
\draw[red,thick,dashed] (-4.5,0)--(-2.5,0) -- (-2.5,1) -- (2.5,1) -- (2.5,0) -- (4.5,0);
\draw[thick, fill=red,red] (-4,0) circle (0.04cm);
\draw[thick, fill=red,red] (-3,0) circle (0.04cm);
\draw[thick, fill=red,red] (-2,1) circle (0.04cm);
\draw[thick, fill=red,red] (-1,1) circle (0.04cm);
\draw[thick, fill=red,red] (0,1) circle (0.04cm);
\draw[thick, fill=red,red] (1,1) circle (0.04cm);
\draw[thick, fill=red,red] (2,1) circle (0.04cm);
\draw[thick, fill=red,red] (3,0) circle (0.04cm);
\draw[thick, fill=red,red] (4,0) circle (0.04cm);
\draw[blue,thick,dashed] (0,2.7) to [out=0,in=125] (1.5,1) to [out=300,in=178] (4.5,0.1);
\draw[blue,thick,dashed] (0,2.7) to [out=180,in=55] (-1.5,1) to [out=240,in=2] (-4.5,0.1);
\draw[thick, fill=blue,blue] (-4,0.1) circle (0.04cm);
\draw[thick, fill=blue,blue] (-3,0.15) circle (0.04cm);
\draw[thick, fill=blue,blue] (-2,0.44) circle (0.04cm);
\draw[thick, fill=blue,blue] (-1,1.93) circle (0.04cm);
\draw[thick, fill=blue,blue] (0,2.7) circle (0.04cm);
\draw[thick, fill=blue,blue] (1,1.93) circle (0.04cm);
\draw[thick, fill=blue,blue] (2,0.44) circle (0.04cm);
\draw[thick, fill=blue,blue] (3,0.15) circle (0.04cm);
\draw[thick, fill=blue,blue] (4,0.1) circle (0.04cm);
\draw[green!50!black,thick,dashed] (0,3.7) to [out=280,in=115] (0.8,1.1) to [out=300,in=179] (4.5,0.1);
\draw[green!50!black,thick,dashed] (0,3.7) to [out=260,in=65] (-0.8,1.1) to [out=240,in=1] (-4.5,0.1);
\draw[thick, fill=green!50!black,green!50!black] (-4,0.1) circle (0.04cm);
\draw[thick, fill=green!50!black,green!50!black] (-3,0.1) circle (0.04cm);
\draw[thick, fill=green!50!black,green!50!black] (-2,0.2) circle (0.04cm);
\draw[thick, fill=green!50!black,green!50!black] (-1,0.8) circle (0.04cm);
\draw[thick, fill=green!50!black,green!50!black] (0,3.7) circle (0.04cm);
\draw[thick, fill=green!50!black,green!50!black] (1,0.8) circle (0.04cm);
\draw[thick, fill=green!50!black,green!50!black] (2,0.2) circle (0.04cm);
\draw[thick, fill=green!50!black,green!50!black] (3,0.1) circle (0.04cm);
\draw[thick, fill=green!50!black,green!50!black] (4,0.1) circle (0.04cm);
\end{tikzpicture}
}}
\caption{Examples for the schematic forms of the function $F$ in $d=1$.}
\label{Fig:SchematicFunctions}
\end{center}
\end{figure}

\noindent
Apart from (\ref{SIConditionFunction}), we pose no further restrictions on $F$. Intuitively, however, to model an epidemic, we expect $F(\vec{X}-\vec{X}')$ to be peaked around $\vec{X}$, different examples of which are schematically shown in Figure~\ref{Fig:SchematicFunctions} for the case $d=1$.\footnote{Notice that we consider in general $F(0)\neq 0$: indeed, $\mathfrak{I}$ models the number of infectious individuals at $\vec{X}$ which are also capable of infecting susceptible individuals at the same lattice site. This is in contrast to the functions $F$ that we had used in the examples in Section~\ref{Sect:StochasticExamples} to model the probabilities for an individual at $\vec{X}$ to become infected by their neighbours.} While each one of these functions leads to a different spread of the disease across the lattice (\emph{i.e.} a qualitatively and quantitatively different solution of (\ref{EqDiffSimpleEpidemic})), we shall define a family of effective (\emph{i.e.} coarse grained) theories, in which the detailed shape of this function becomes parametrically less and less important.

\subsubsection{Coarse Graining: Continuous Spatial Variables}\label{Sect:SICoarseGraining}
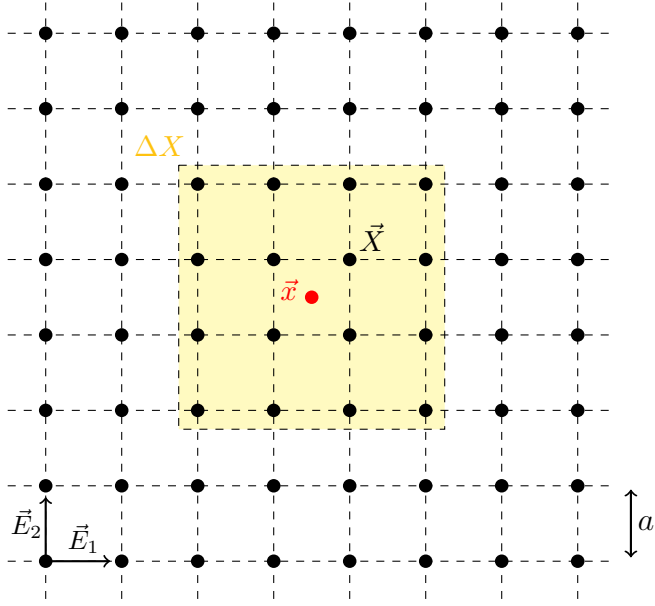
\begin{figure}[htbp]
\begin{center}
\scalebox{1}{\parbox{8.7cm}{\begin{tikzpicture}
\draw[dashed, fill=yellow!30!white] (-2.25,1.25) -- (1.25,1.25) -- (1.25,-2.25) -- (-2.25,-2.25) -- (-2.25,1.25);
%
\draw[thick, fill=black] (-4,3) circle (0.075cm);
\draw[thick, fill=black] (-3,3) circle (0.075cm);
\draw[thick, fill=black] (-2,3) circle (0.075cm);
\draw[thick, fill=black] (-1,3) circle (0.075cm);
\draw[thick, fill=black] (0,3) circle (0.075cm);
\draw[thick, fill=black] (1,3) circle (0.075cm);
\draw[thick, fill=black] (2,3) circle (0.075cm);
\draw[thick, fill=black] (3,3) circle (0.075cm);
%
\draw[thick, fill=black] (-4,2) circle (0.075cm);
\draw[thick, fill=black] (-3,2) circle (0.075cm);
\draw[thick, fill=black] (-2,2) circle (0.075cm);
\draw[thick, fill=black] (-1,2) circle (0.075cm);
\draw[thick, fill=black] (0,2) circle (0.075cm);
\draw[thick, fill=black] (1,2) circle (0.075cm);
\draw[thick, fill=black] (2,2) circle (0.075cm);
\draw[thick, fill=black] (3,2) circle (0.075cm);
%
\draw[thick, fill=black] (-4,1) circle (0.075cm);
\draw[thick, fill=black] (-3,1) circle (0.075cm);
\draw[thick, fill=black] (-2,1) circle (0.075cm);
\draw[thick, fill=black] (-1,1) circle (0.075cm);
\draw[thick, fill=black] (0,1) circle (0.075cm);
\draw[thick, fill=black] (1,1) circle (0.075cm);
\draw[thick, fill=black] (2,1) circle (0.075cm);
\draw[thick, fill=black] (3,1) circle (0.075cm);
%
\draw[thick, fill=black] (-4,0) circle (0.075cm);
\draw[thick, fill=black] (-3,0) circle (0.075cm);
\draw[thick, fill=black] (-2,0) circle (0.075cm);
\draw[thick, fill=black] (-1,0) circle (0.075cm);
\draw[thick, fill=black] (0,0) circle (0.075cm);
\draw[thick, fill=black] (1,0) circle (0.075cm);
\draw[thick, fill=black] (2,0) circle (0.075cm);
\draw[thick, fill=black] (3,0) circle (0.075cm);
%
\draw[thick, fill=black] (-4,-1) circle (0.075cm);
\draw[thick, fill=black] (-3,-1) circle (0.075cm);
\draw[thick, fill=black] (-2,-1) circle (0.075cm);
\draw[thick, fill=black] (-1,-1) circle (0.075cm);
\draw[thick, fill=black] (0,-1) circle (0.075cm);
\draw[thick, fill=black] (1,-1) circle (0.075cm);
\draw[thick, fill=black] (2,-1) circle (0.075cm);
\draw[thick, fill=black] (3,-1) circle (0.075cm);
%
\draw[thick, fill=black] (-4,-2) circle (0.075cm);
\draw[thick, fill=black] (-3,-2) circle (0.075cm);
\draw[thick, fill=black] (-2,-2) circle (0.075cm);
\draw[thick, fill=black] (-1,-2) circle (0.075cm);
\draw[thick, fill=black] (0,-2) circle (0.075cm);
\draw[thick, fill=black] (1,-2) circle (0.075cm);
\draw[thick, fill=black] (2,-2) circle (0.075cm);
\draw[thick, fill=black] (3,-2) circle (0.075cm);
%
\draw[thick, fill=black] (-4,-3) circle (0.075cm);
\draw[thick, fill=black] (-3,-3) circle (0.075cm);
\draw[thick, fill=black] (-2,-3) circle (0.075cm);
\draw[thick, fill=black] (-1,-3) circle (0.075cm);
\draw[thick, fill=black] (0,-3) circle (0.075cm);
\draw[thick, fill=black] (1,-3) circle (0.075cm);
\draw[thick, fill=black] (2,-3) circle (0.075cm);
\draw[thick, fill=black] (3,-3) circle (0.075cm);
%
\draw[thick, fill=black] (-4,-4) circle (0.075cm);
\draw[thick, fill=black] (-3,-4) circle (0.075cm);
\draw[thick, fill=black] (-2,-4) circle (0.075cm);
\draw[thick, fill=black] (-1,-4) circle (0.075cm);
\draw[thick, fill=black] (0,-4) circle (0.075cm);
\draw[thick, fill=black] (1,-4) circle (0.075cm);
\draw[thick, fill=black] (2,-4) circle (0.075cm);
\draw[thick, fill=black] (3,-4) circle (0.075cm);
%
\draw[<->,thick] (3.7,-3.95) -- (3.7,-3.05);
\node at (3.9,.-3.5) {$a$};
%
\draw[thick, fill=black,red] (-0.5,-0.5) circle (0.075cm);
\node[red] at (-0.8,-0.4) {\small $\vec{x}$};
\node[yellow!75!red] at (-2.5,1.5) {\small $\Delta X$};
\node at (0.3,0.3) {\small $\vec{X}$};
%
\draw[dashed] (-4.5,3) -- (3.5,3);
\draw[dashed] (-4.5,2) -- (3.5,2);
\draw[dashed] (-4.5,1) -- (3.5,1);
\draw[dashed] (-4.5,0) -- (3.5,0);
\draw[dashed] (-4.5,-1) -- (3.5,-1);
\draw[dashed] (-4.5,-2) -- (3.5,-2);
\draw[dashed] (-4.5,-3) -- (3.5,-3);
\draw[dashed] (-4.5,-4) -- (3.5,-4);
%
\draw[dashed] (3,-4.5) -- (3,3.5);
\draw[dashed] (2,-4.5) -- (2,3.5);
\draw[dashed] (1,-4.5) -- (1,3.5);
\draw[dashed] (0,-4.5) -- (0,3.5);
\draw[dashed] (-1,-4.5) -- (-1,3.5);
\draw[dashed] (-2,-4.5) -- (-2,3.5);
\draw[dashed] (-3,-4.5) -- (-3,3.5);
\draw[dashed] (-4,-4.5) -- (-4,3.5);
\draw[thick,->] (-4,-4) -- (-3.15,-4);
\node at (-3.5,-3.7) {\footnotesize $\vec{E}_1$};
\draw[thick,->] (-4,-4) -- (-4,-3.15);
\node at (-4.25,-3.5) {\footnotesize $\vec{E}_2$};
\end{tikzpicture}
}}
\caption{Coarse graining in the case of epidemical model~(\ref{EqDiffSimpleEpidemic}).}
\label{Fig:CoarseGrainingSI}
\end{center}
\end{figure}

Following the example of the diffusion process in Section~\ref{Sect:CoarseGrainingDiffusion}, we shall now pass to continuous space-variables and approximate (\ref{EqDiffSimpleEpidemic}) by a partial differential equation for continuous variables $\vec{x}\in\mathbb{R}$ instead of discrete $\vec{X}\in\Gamma$. To this end, instead of the number of infectious individuals at lattice site $\vec{X}\in\Gamma$ we introduce the coarse grained quantity $I:\,\mathbb{R}^d\times \mathbb{R}_+\longrightarrow \mathbb{R}_+$ as the average of the number of infectious individuals over a compact subset $\Delta X\subset \Gamma$ 
\begin{align}
&\mathfrak{I}(\vec{X},t)\longrightarrow I(\vec{x},t)=\frac{1}{|\Delta X|}\sum_{\vec{X}\,'\in\Delta X}\mathfrak{I}(\vec{X}\,',t)\sim \mathfrak{I}(\vec{X},t)\,,&&\text{with} &&\vec{x}\in\mathbb{R}^d\,,\label{CoarseGrainingSIinf}
\end{align}
where $|\Delta X|$ is the number of lattice points contained in $\Delta X$.\footnote{We note that one could introduce a constant factor in eq.~(\ref{CoarseGrainingSIinf}), which reflects the liberty of rescaling of the (normalised) $I$. For the sake of simplicity, we refrain from doing so here. Similarly, we note that we could also define $|\Delta X|=\text{vol}_{\mathbb{R}^d}(\Delta X)/a^d$, which is equivalent up to a rescaling of $I(\vec{x},t)$. We shall comment on the effect of the latter below.} The region $\Delta X$ is labeled by a continuous variable $\vec{x}\in\mathbb{R}^d$ (where potentially $\vec{x}\notin\Gamma$), as is schematically shown in Figure~\ref{Fig:CoarseGrainingSI}. In the last step in (\ref{CoarseGrainingSIinf}) we have assumed that the number of infectious individuals does not vary significantly over $\Delta X$, such that the value of $I(\vec{x},t)$ can be identified with $\mathfrak{I}(\vec{X},t)$. We furthermore assume that $I(\vec{x},t)$ is a continuous function, such that in (\ref{EqDiffSimpleEpidemic}) we can replace

\noindent
\begin{align}
\sum_{\vec{X}\,'\in\Gamma}\,F(\vec{X}-\vec{X}\,')\,\mathfrak{I}(\vec{X}\,',t)\longrightarrow \int_{\mathbb{R}^d} d^dx'\,f(\vec{x}-\vec{x}\,')\,I(\vec{x}\,',t)=(f* I)(\vec{x},t)\,,\label{ConvolutionTerm}
\end{align}
where $*$ denotes the convolution (with respect to the spatial variable $\vec{x}$) and $f:\,\mathbb{R}^d\longrightarrow \mathbb{R}_+$ is a continuous function which satisfies
\begin{align}
&\int_{\mathbb{R}^d} d^dx\,f(\vec{x})=\gamma\,,&&\text{and} &&f(\vec{X})=F(\vec{X})\,,\hspace{0.5cm}\forall \vec{X}\in\Gamma\cap\mathbb{R}^d\,,\label{PropFuncF}
\end{align}
for some real constant $\gamma$ with $0<\gamma<\infty$.\footnote{The relation between $\gamma$ and $\gamma'$ in (\ref{SIConditionFunction}) depends on the details how $F$ is continued from a function defined on $\Gamma$ to a function defined on all $\mathbb{R}^d$, which shall not be important in the following.} Notice, since $f$ is assumed to be non-negative for all $\vec{x}\in\mathbb{R}^d$, the first condition implies that $f\in L^1(\mathbb{R}^d)$. We therefore write the following coarse grained approximation of eq.(\ref{EqDiffSimpleEpidemic})
\begin{align}
\frac{\partial I}{\partial t}(\vec{x},t)=\left(1-I(\vec{x},t)\right)\,\left((f*I)(\vec{x},t)\right)\,.\label{EqDiffSimpleEpidemicRed}
\end{align}
This model has been studied in \cite{Mollison1977} for $d=1$, in particular searching for travelling waves and relating it to stochastic approaches~\cite{MollisonVelocities}. 
\subsection{Family of Effective Models}\label{Sect:SIeffectivemodels}
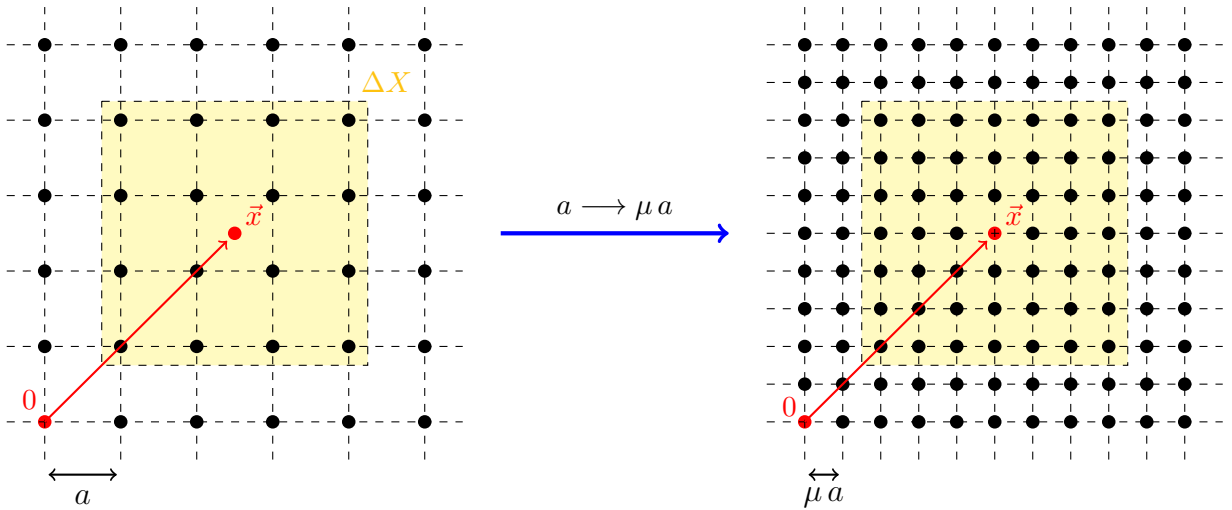
\begin{figure}[htbp]
\begin{center}
\scalebox{1}{\parbox{16.1cm}{\begin{tikzpicture}
\draw[dashed, fill=yellow!30!white] (-2.25,1.25) -- (1.25,1.25) -- (1.25,-2.25) -- (-2.25,-2.25) -- (-2.25,1.25);
\draw[thick, fill=black] (-3,2) circle (0.075cm);
\draw[thick, fill=black] (-2,2) circle (0.075cm);
\draw[thick, fill=black] (-1,2) circle (0.075cm);
\draw[thick, fill=black] (0,2) circle (0.075cm);
\draw[thick, fill=black] (1,2) circle (0.075cm);
\draw[thick, fill=black] (2,2) circle (0.075cm);
\draw[thick, fill=black] (-3,1) circle (0.075cm);
\draw[thick, fill=black] (-2,1) circle (0.075cm);
\draw[thick, fill=black] (-1,1) circle (0.075cm);
\draw[thick, fill=black] (0,1) circle (0.075cm);
\draw[thick, fill=black] (1,1) circle (0.075cm);
\draw[thick, fill=black] (2,1) circle (0.075cm);
\draw[thick, fill=black] (-3,0) circle (0.075cm);
\draw[thick, fill=black] (-2,0) circle (0.075cm);
\draw[thick, fill=black] (-1,0) circle (0.075cm);
\draw[thick, fill=black] (0,0) circle (0.075cm);
\draw[thick, fill=black] (1,0) circle (0.075cm);
\draw[thick, fill=black] (2,0) circle (0.075cm);
\draw[thick, fill=black] (-3,-1) circle (0.075cm);
\draw[thick, fill=black] (-2,-1) circle (0.075cm);
\draw[thick, fill=black] (-1,-1) circle (0.075cm);
\draw[thick, fill=black] (0,-1) circle (0.075cm);
\draw[thick, fill=black] (1,-1) circle (0.075cm);
\draw[thick, fill=black] (2,-1) circle (0.075cm);
\draw[thick, fill=black] (-3,-2) circle (0.075cm);
\draw[thick, fill=black] (-2,-2) circle (0.075cm);
\draw[thick, fill=black] (-1,-2) circle (0.075cm);
\draw[thick, fill=black] (0,-2) circle (0.075cm);
\draw[thick, fill=black] (1,-2) circle (0.075cm);
\draw[thick, fill=black] (2,-2) circle (0.075cm);
\draw[thick, fill=black,red] (-3,-3) circle (0.075cm);
\draw[thick, fill=black] (-2,-3) circle (0.075cm);
\draw[thick, fill=black] (-1,-3) circle (0.075cm);
\draw[thick, fill=black] (0,-3) circle (0.075cm);
\draw[thick, fill=black] (1,-3) circle (0.075cm);
\draw[thick, fill=black] (2,-3) circle (0.075cm);
\draw[<->,thick] (-2.95,-3.7) -- (-2.05,-3.7);
\node at (-2.5,-4) {$a$};
\node[yellow!75!red] at (1.5,1.5) {\small $\Delta X$};
\draw[dashed] (-3.5,2) -- (2.5,2);
\draw[dashed] (-3.5,1) -- (2.5,1);
\draw[dashed] (-3.5,0) -- (2.5,0);
\draw[dashed] (-3.5,-1) -- (2.5,-1);
\draw[dashed] (-3.5,-2) -- (2.5,-2);
\draw[dashed] (-3.5,-3) -- (2.5,-3);
\draw[dashed] (2,-3.5) -- (2,2.5);
\draw[dashed] (1,-3.5) -- (1,2.5);
\draw[dashed] (0,-3.5) -- (0,2.5);
\draw[dashed] (-1,-3.5) -- (-1,2.5);
\draw[dashed] (-2,-3.5) -- (-2,2.5);
\draw[dashed] (-3,-3.5) -- (-3,2.5);
\draw[thick, fill=black,red] (-0.5,-0.5) circle (0.075cm);
\node[red] at (-3.2,-2.7) {\small $0$};
\draw[thick,->,red] (-3,-3) -- (-0.6,-0.6);
\node[red] at (-0.25,-0.25) {\small $\vec{x}$};
\draw[ultra thick,->,blue] (3,-0.5) -- (6,-0.5);
\node at (4.5,-0.2) {\small $a\longrightarrow \mu\,a$};
\begin{scope}[xshift=10cm]
\draw[dashed, fill=yellow!30!white] (-2.25,1.25) -- (1.25,1.25) -- (1.25,-2.25) -- (-2.25,-2.25) -- (-2.25,1.25);
\draw[thick, fill=black,red] (-3,-3) circle (0.075cm);
\draw[thick, fill=black] (-2.5,-3) circle (0.075cm);
\draw[thick, fill=black] (-2,-3) circle (0.075cm);
\draw[thick, fill=black] (-1.5,-3) circle (0.075cm);
\draw[thick, fill=black] (-1,-3) circle (0.075cm);
\draw[thick, fill=black] (-0.5,-3) circle (0.075cm);
\draw[thick, fill=black] (0,-3) circle (0.075cm);
\draw[thick, fill=black] (0.5,-3) circle (0.075cm);
\draw[thick, fill=black] (1,-3) circle (0.075cm);
\draw[thick, fill=black] (1.5,-3) circle (0.075cm);
\draw[thick, fill=black] (2,-3) circle (0.075cm);
\draw[thick, fill=black] (-3,-2.5) circle (0.075cm);
\draw[thick, fill=black] (-2.5,-2.5) circle (0.075cm);
\draw[thick, fill=black] (-2,-2.5) circle (0.075cm);
\draw[thick, fill=black] (-1.5,-2.5) circle (0.075cm);
\draw[thick, fill=black] (-1,-2.5) circle (0.075cm);
\draw[thick, fill=black] (-0.5,-2.5) circle (0.075cm);
\draw[thick, fill=black] (0,-2.5) circle (0.075cm);
\draw[thick, fill=black] (0.5,-2.5) circle (0.075cm);
\draw[thick, fill=black] (1,-2.5) circle (0.075cm);
\draw[thick, fill=black] (1.5,-2.5) circle (0.075cm);
\draw[thick, fill=black] (2,-2.5) circle (0.075cm);
\draw[thick, fill=black] (-3,-2) circle (0.075cm);
\draw[thick, fill=black] (-2.5,-2) circle (0.075cm);
\draw[thick, fill=black] (-2,-2) circle (0.075cm);
\draw[thick, fill=black] (-1.5,-2) circle (0.075cm);
\draw[thick, fill=black] (-1,-2) circle (0.075cm);
\draw[thick, fill=black] (-0.5,-2) circle (0.075cm);
\draw[thick, fill=black] (0,-2) circle (0.075cm);
\draw[thick, fill=black] (0.5,-2) circle (0.075cm);
\draw[thick, fill=black] (1,-2) circle (0.075cm);
\draw[thick, fill=black] (1.5,-2) circle (0.075cm);
\draw[thick, fill=black] (2,-2) circle (0.075cm);
\draw[thick, fill=black] (-3,-1.5) circle (0.075cm);
\draw[thick, fill=black] (-2.5,-1.5) circle (0.075cm);
\draw[thick, fill=black] (-2,-1.5) circle (0.075cm);
\draw[thick, fill=black] (-1.5,-1.5) circle (0.075cm);
\draw[thick, fill=black] (-1,-1.5) circle (0.075cm);
\draw[thick, fill=black] (-0.5,-1.5) circle (0.075cm);
\draw[thick, fill=black] (0,-1.5) circle (0.075cm);
\draw[thick, fill=black] (0.5,-1.5) circle (0.075cm);
\draw[thick, fill=black] (1,-1.5) circle (0.075cm);
\draw[thick, fill=black] (1.5,-1.5) circle (0.075cm);
\draw[thick, fill=black] (2,-1.5) circle (0.075cm);
\draw[thick, fill=black] (-3,-1) circle (0.075cm);
\draw[thick, fill=black] (-2.5,-1) circle (0.075cm);
\draw[thick, fill=black] (-2,-1) circle (0.075cm);
\draw[thick, fill=black] (-1.5,-1) circle (0.075cm);
\draw[thick, fill=black] (-1,-1) circle (0.075cm);
\draw[thick, fill=black] (-0.5,-1) circle (0.075cm);
\draw[thick, fill=black] (0,-1) circle (0.075cm);
\draw[thick, fill=black] (0.5,-1) circle (0.075cm);
\draw[thick, fill=black] (1,-1) circle (0.075cm);
\draw[thick, fill=black] (1.5,-1) circle (0.075cm);
\draw[thick, fill=black] (2,-1) circle (0.075cm);
\draw[thick, fill=black] (-3,-0.5) circle (0.075cm);
\draw[thick, fill=black] (-2.5,-0.5) circle (0.075cm);
\draw[thick, fill=black] (-2,-0.5) circle (0.075cm);
\draw[thick, fill=black] (-1.5,-0.5) circle (0.075cm);
\draw[thick, fill=black] (-1,-0.5) circle (0.075cm);
\draw[thick, fill=black,red] (-0.5,-0.5) circle (0.075cm);
\draw[thick, fill=black] (0,-0.5) circle (0.075cm);
\draw[thick, fill=black] (0.5,-0.5) circle (0.075cm);
\draw[thick, fill=black] (1,-0.5) circle (0.075cm);
\draw[thick, fill=black] (1.5,-0.5) circle (0.075cm);
\draw[thick, fill=black] (2,-0.5) circle (0.075cm);
\draw[thick, fill=black] (-3,0) circle (0.075cm);
\draw[thick, fill=black] (-2.5,0) circle (0.075cm);
\draw[thick, fill=black] (-2,0) circle (0.075cm);
\draw[thick, fill=black] (-1.5,0) circle (0.075cm);
\draw[thick, fill=black] (-1,0) circle (0.075cm);
\draw[thick, fill=black] (-0.5,0) circle (0.075cm);
\draw[thick, fill=black] (0,0) circle (0.075cm);
\draw[thick, fill=black] (0.5,0) circle (0.075cm);
\draw[thick, fill=black] (1,0) circle (0.075cm);
\draw[thick, fill=black] (1.5,0) circle (0.075cm);
\draw[thick, fill=black] (2,0) circle (0.075cm);
\draw[thick, fill=black] (-3,0.5) circle (0.075cm);
\draw[thick, fill=black] (-2.5,0.5) circle (0.075cm);
\draw[thick, fill=black] (-2,0.5) circle (0.075cm);
\draw[thick, fill=black] (-1.5,0.5) circle (0.075cm);
\draw[thick, fill=black] (-1,0.5) circle (0.075cm);
\draw[thick, fill=black] (-0.5,0.5) circle (0.075cm);
\draw[thick, fill=black] (0,0.5) circle (0.075cm);
\draw[thick, fill=black] (0.5,0.5) circle (0.075cm);
\draw[thick, fill=black] (1,0.5) circle (0.075cm);
\draw[thick, fill=black] (1.5,0.5) circle (0.075cm);
\draw[thick, fill=black] (2,0.5) circle (0.075cm);
\draw[thick, fill=black] (-3,1) circle (0.075cm);
\draw[thick, fill=black] (-2.5,1) circle (0.075cm);
\draw[thick, fill=black] (-2,1) circle (0.075cm);
\draw[thick, fill=black] (-1.5,1) circle (0.075cm);
\draw[thick, fill=black] (-1,1) circle (0.075cm);
\draw[thick, fill=black] (-0.5,1) circle (0.075cm);
\draw[thick, fill=black] (0,1) circle (0.075cm);
\draw[thick, fill=black] (0.5,1) circle (0.075cm);
\draw[thick, fill=black] (1,1) circle (0.075cm);
\draw[thick, fill=black] (1.5,1) circle (0.075cm);
\draw[thick, fill=black] (2,1) circle (0.075cm);
\draw[thick, fill=black] (-3,1.5) circle (0.075cm);
\draw[thick, fill=black] (-2.5,1.5) circle (0.075cm);
\draw[thick, fill=black] (-2,1.5) circle (0.075cm);
\draw[thick, fill=black] (-1.5,1.5) circle (0.075cm);
\draw[thick, fill=black] (-1,1.5) circle (0.075cm);
\draw[thick, fill=black] (-0.5,1.5) circle (0.075cm);
\draw[thick, fill=black] (0,1.5) circle (0.075cm);
\draw[thick, fill=black] (0.5,1.5) circle (0.075cm);
\draw[thick, fill=black] (1,1.5) circle (0.075cm);
\draw[thick, fill=black] (1.5,1.5) circle (0.075cm);
\draw[thick, fill=black] (2,1.5) circle (0.075cm);
\draw[thick, fill=black] (-3,2) circle (0.075cm);
\draw[thick, fill=black] (-2.5,2) circle (0.075cm);
\draw[thick, fill=black] (-2,2) circle (0.075cm);
\draw[thick, fill=black] (-1.5,2) circle (0.075cm);
\draw[thick, fill=black] (-1,2) circle (0.075cm);
\draw[thick, fill=black] (-0.5,2) circle (0.075cm);
\draw[thick, fill=black] (0,2) circle (0.075cm);
\draw[thick, fill=black] (0.5,2) circle (0.075cm);
\draw[thick, fill=black] (1,2) circle (0.075cm);
\draw[thick, fill=black] (1.5,2) circle (0.075cm);
\draw[thick, fill=black] (2,2) circle (0.075cm);
\draw[dashed] (-3.5,2) -- (2.5,2);
\draw[dashed] (-3.5,1.5) -- (2.5,1.5);
\draw[dashed] (-3.5,1) -- (2.5,1);
\draw[dashed] (-3.5,0.5) -- (2.5,0.5);
\draw[dashed] (-3.5,0) -- (2.5,0);
\draw[dashed] (-3.5,-0.5) -- (2.5,-0.5);
\draw[dashed] (-3.5,-1) -- (2.5,-1);
\draw[dashed] (-3.5,-1.5) -- (2.5,-1.5);
\draw[dashed] (-3.5,-2) -- (2.5,-2);
\draw[dashed] (-3.5,-2.5) -- (2.5,-2.5);
\draw[dashed] (-3.5,-3) -- (2.5,-3);
\draw[dashed] (-3,-3.5) -- (-3,2.5);
\draw[dashed] (-2.5,-3.5) -- (-2.5,2.5);
\draw[dashed] (-2,-3.5) -- (-2,2.5);
\draw[dashed] (-1.5,-3.5) -- (-1.5,2.5);
\draw[dashed] (-1,-3.5) -- (-1,2.5);
\draw[dashed] (-0.5,-3.5) -- (-0.5,2.5);
\draw[dashed] (0,-3.5) -- (0,2.5);
\draw[dashed] (0.5,-3.5) -- (0.5,2.5);
\draw[dashed] (1,-3.5) -- (1,2.5);
\draw[dashed] (1.5,-3.5) -- (1.5,2.5);
\draw[dashed] (2,-3.5) -- (2,2.5);
\draw[<->,thick] (-2.95,-3.7) -- (-2.55,-3.7);
\node at (-2.75,-4) {$\mu\,a$};
\node[red] at (-3.2,-2.8) {\small $0$};
\draw[thick,->,red] (-3,-3) -- (-0.6,-0.6);
\node[red] at (-0.25,-0.25) {\small $\vec{x}$};
\end{scope}
\end{tikzpicture}
}}
\caption{Rescaling of $\Gamma$ for $0<\mu<1$: both $\Delta X\subset\mathbb{R}^d$ and $\vec{x}\in\mathbb{R}^d$ remain fixed.}
\label{Fig:EffectiveTheoriesRescaling1}
\end{center}
\end{figure}

The equation (\ref{EqDiffSimpleEpidemicRed}) provides an effective description of a (microscopic) model based on a lattice $\Gamma$ with lattice spacing $a$ through coarse graining. In the following, we shall consider a family of such effective theories, where we rescale $a$ by a dimensionless parameter $\mu$
\begin{align}
&a\longrightarrow \mu\,a\,,&&\text{with} &&\mu\in\mathbb{R}_+\,.\label{RescalLatt}
\end{align}
We shall, however, keep $\Delta X$ as a subset of $\mathbb{R}^d$ fixed, as is schematically shown in Figure~\ref{Fig:EffectiveTheoriesRescaling1} for $0<\mu<1$.\footnote{In the following we shall be mostly interested in the limit $\mu\to 0$, such that implicitly (\ref{RescalLatt}) corresponds to considering more fine-grained lattices.} A possible way to interpret this from an epidemiological perspective, is to associate $\Delta X$ (labelled by the fixed $\vec{x}\in\mathbb{R}^d$) with a (fixed) geographical region, while the lattice points represent (finer and finer) subdivisions of the population (\emph{e.g.} provinces, cities, households, \emph{etc.}). Concretely one may think of $\Delta X$ as a country with the lattice points representing its different provinces. After rescaling with a certain $\mu$, we obtain a new model in which the spread of the disease within the same country $\Delta X$ is now described at the level of cities (represented by the finer grained lattice).


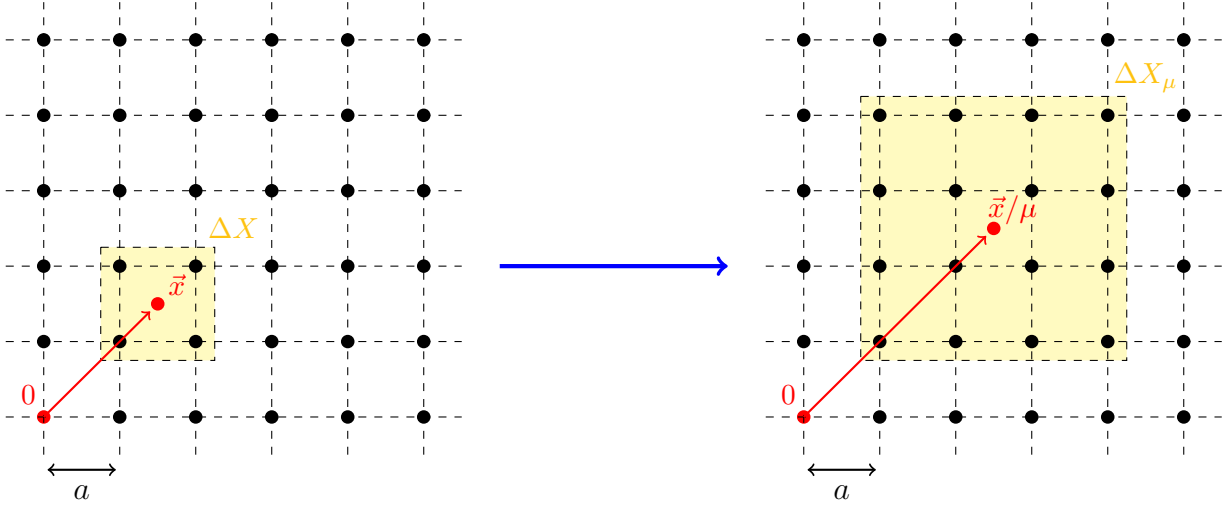
\begin{figure}[htbp]
\begin{center}
\scalebox{1}{\parbox{16.1cm}{\begin{tikzpicture}
\draw[dashed, fill=yellow!30!white] (-2.25,-0.75) -- (-0.75,-0.75) -- (-0.75,-2.25) -- (-2.25,-2.25) -- (-2.25,-0.75);
\draw[thick, fill=black] (-3,2) circle (0.075cm);
\draw[thick, fill=black] (-2,2) circle (0.075cm);
\draw[thick, fill=black] (-1,2) circle (0.075cm);
\draw[thick, fill=black] (0,2) circle (0.075cm);
\draw[thick, fill=black] (1,2) circle (0.075cm);
\draw[thick, fill=black] (2,2) circle (0.075cm);
\draw[thick, fill=black] (-3,1) circle (0.075cm);
\draw[thick, fill=black] (-2,1) circle (0.075cm);
\draw[thick, fill=black] (-1,1) circle (0.075cm);
\draw[thick, fill=black] (0,1) circle (0.075cm);
\draw[thick, fill=black] (1,1) circle (0.075cm);
\draw[thick, fill=black] (2,1) circle (0.075cm);
\draw[thick, fill=black] (-3,0) circle (0.075cm);
\draw[thick, fill=black] (-2,0) circle (0.075cm);
\draw[thick, fill=black] (-1,0) circle (0.075cm);
\draw[thick, fill=black] (0,0) circle (0.075cm);
\draw[thick, fill=black] (1,0) circle (0.075cm);
\draw[thick, fill=black] (2,0) circle (0.075cm);
\draw[thick, fill=black] (-3,-1) circle (0.075cm);
\draw[thick, fill=black] (-2,-1) circle (0.075cm);
\draw[thick, fill=black] (-1,-1) circle (0.075cm);
\draw[thick, fill=black] (0,-1) circle (0.075cm);
\draw[thick, fill=black] (1,-1) circle (0.075cm);
\draw[thick, fill=black] (2,-1) circle (0.075cm);
\draw[thick, fill=black] (-3,-2) circle (0.075cm);
\draw[thick, fill=black] (-2,-2) circle (0.075cm);
\draw[thick, fill=black] (-1,-2) circle (0.075cm);
\draw[thick, fill=black] (0,-2) circle (0.075cm);
\draw[thick, fill=black] (1,-2) circle (0.075cm);
\draw[thick, fill=black] (2,-2) circle (0.075cm);
\draw[thick, fill=black,red] (-3,-3) circle (0.075cm);
\draw[thick, fill=black] (-2,-3) circle (0.075cm);
\draw[thick, fill=black] (-1,-3) circle (0.075cm);
\draw[thick, fill=black] (0,-3) circle (0.075cm);
\draw[thick, fill=black] (1,-3) circle (0.075cm);
\draw[thick, fill=black] (2,-3) circle (0.075cm);
\draw[<->,thick] (-2.95,-3.7) -- (-2.05,-3.7);
\node at (-2.5,-4) {$a$};
\node[yellow!75!red] at (-0.5,-0.5) {\small $\Delta X$};
\draw[dashed] (-3.5,2) -- (2.5,2);
\draw[dashed] (-3.5,1) -- (2.5,1);
\draw[dashed] (-3.5,0) -- (2.5,0);
\draw[dashed] (-3.5,-1) -- (2.5,-1);
\draw[dashed] (-3.5,-2) -- (2.5,-2);
\draw[dashed] (-3.5,-3) -- (2.5,-3);
\draw[dashed] (2,-3.5) -- (2,2.5);
\draw[dashed] (1,-3.5) -- (1,2.5);
\draw[dashed] (0,-3.5) -- (0,2.5);
\draw[dashed] (-1,-3.5) -- (-1,2.5);
\draw[dashed] (-2,-3.5) -- (-2,2.5);
\draw[dashed] (-3,-3.5) -- (-3,2.5);
\draw[thick, fill=black,red] (-1.5,-1.5) circle (0.075cm);
\node[red] at (-3.2,-2.7) {\small $0$};
\draw[thick,->,red] (-3,-3) -- (-1.6,-1.6);
\node[red] at (-1.25,-1.25) {\small $\vec{x}$};
%
%
\draw[ultra thick,->,blue] (3,-1) -- (6,-1);
\begin{scope}[xshift=10cm]
\draw[dashed, fill=yellow!30!white] (-2.25,1.25) -- (1.25,1.25) -- (1.25,-2.25) -- (-2.25,-2.25) -- (-2.25,1.25);
\draw[thick, fill=black] (-3,2) circle (0.075cm);
\draw[thick, fill=black] (-2,2) circle (0.075cm);
\draw[thick, fill=black] (-1,2) circle (0.075cm);
\draw[thick, fill=black] (0,2) circle (0.075cm);
\draw[thick, fill=black] (1,2) circle (0.075cm);
\draw[thick, fill=black] (2,2) circle (0.075cm);
\draw[thick, fill=black] (-3,1) circle (0.075cm);
\draw[thick, fill=black] (-2,1) circle (0.075cm);
\draw[thick, fill=black] (-1,1) circle (0.075cm);
\draw[thick, fill=black] (0,1) circle (0.075cm);
\draw[thick, fill=black] (1,1) circle (0.075cm);
\draw[thick, fill=black] (2,1) circle (0.075cm);
\draw[thick, fill=black] (-3,0) circle (0.075cm);
\draw[thick, fill=black] (-2,0) circle (0.075cm);
\draw[thick, fill=black] (-1,0) circle (0.075cm);
\draw[thick, fill=black] (0,0) circle (0.075cm);
\draw[thick, fill=black] (1,0) circle (0.075cm);
\draw[thick, fill=black] (2,0) circle (0.075cm);
\draw[thick, fill=black] (-3,-1) circle (0.075cm);
\draw[thick, fill=black] (-2,-1) circle (0.075cm);
\draw[thick, fill=black] (-1,-1) circle (0.075cm);
\draw[thick, fill=black] (0,-1) circle (0.075cm);
\draw[thick, fill=black] (1,-1) circle (0.075cm);
\draw[thick, fill=black] (2,-1) circle (0.075cm);
\draw[thick, fill=black] (-3,-2) circle (0.075cm);
\draw[thick, fill=black] (-2,-2) circle (0.075cm);
\draw[thick, fill=black] (-1,-2) circle (0.075cm);
\draw[thick, fill=black] (0,-2) circle (0.075cm);
\draw[thick, fill=black] (1,-2) circle (0.075cm);
\draw[thick, fill=black] (2,-2) circle (0.075cm);
\draw[thick, fill=black,red] (-3,-3) circle (0.075cm);
\draw[thick, fill=black] (-2,-3) circle (0.075cm);
\draw[thick, fill=black] (-1,-3) circle (0.075cm);
\draw[thick, fill=black] (0,-3) circle (0.075cm);
\draw[thick, fill=black] (1,-3) circle (0.075cm);
\draw[thick, fill=black] (2,-3) circle (0.075cm);
\draw[<->,thick] (-2.95,-3.7) -- (-2.05,-3.7);
\node at (-2.5,-4) {$a$};
\node[yellow!75!red] at (1.5,1.5) {\small $\Delta X_\mu$};
\draw[dashed] (-3.5,2) -- (2.5,2);
\draw[dashed] (-3.5,1) -- (2.5,1);
\draw[dashed] (-3.5,0) -- (2.5,0);
\draw[dashed] (-3.5,-1) -- (2.5,-1);
\draw[dashed] (-3.5,-2) -- (2.5,-2);
\draw[dashed] (-3.5,-3) -- (2.5,-3);
\draw[dashed] (2,-3.5) -- (2,2.5);
\draw[dashed] (1,-3.5) -- (1,2.5);
\draw[dashed] (0,-3.5) -- (0,2.5);
\draw[dashed] (-1,-3.5) -- (-1,2.5);
\draw[dashed] (-2,-3.5) -- (-2,2.5);
\draw[dashed] (-3,-3.5) -- (-3,2.5);
\draw[thick, fill=black,red] (-0.5,-0.5) circle (0.075cm);
\node[red] at (-3.2,-2.7) {\small $0$};
\draw[thick,->,red] (-3,-3) -- (-0.6,-0.6);
\node[red] at (-0.25,-0.25) {\small $\vec{x}/\mu$};
\end{scope}

\end{tikzpicture}
}}
\caption{Equivalent interpretation of the rescaling for $0<\mu<1$: instead of changing the lattice spacing $a$ of $\Gamma$ the coarse graining is performed over a larger subset $\Delta X_\mu$ of $\Gamma$ and $\vec{x}$ is replaced by $\frac{\vec{x}}{\mu}$.}
\label{Fig:EffectiveTheoriesRescaling2}
\end{center}
\end{figure}

We remark that there is an alternative way to interpret this rescaling from the perspective of the variables $I$ defined in (\ref{CoarseGrainingSIinf}): instead of a change of the lattice spacing (\ref{RescalLatt}), we can think of keeping $a$ fixed, but instead rescaling $\Delta X$ (and the variable $\vec{x}$ labelling it) to include more lattice sites. This is schematically shown in Figure~\ref{Fig:EffectiveTheoriesRescaling2}. From an epidemiological perspective, a smaller value of $\mu$ implies that we are averaging the number of infectious individuals over a larger geographical region: \emph{e.g.} instead of counting the number of infected individuals in a single province (made up from several cities represented by the lattice points), $I$ after rescaling counts the (averaged) number of infected in an entire country (with the lattice points still representing single cities).


Implementing the rescaling (\ref{RescalLatt}) at the level of the equation (\ref{EqDiffSimpleEpidemicRed}), we first realise that, from the perspective of the lattice variables it requires to replace $\mathfrak{I}(\vec{X},t)$ by $\mathfrak{I}(\vec{X}/\mu,t)$ (see Figure~\ref{Fig:EffectiveTheoriesRescaling2}), which therefore suggests to rescale
\begin{align}
I(\vec{x},t)&\longrightarrow I_\mu(\vec{x},t)=I(\vec{x}/\mu,t)\,,\label{ReplaceFunc}
\end{align}
where $I_\mu:\,\mathbb{R}^d\longrightarrow \mathbb{R}_+$ is a regular function. This in turn suggests for the convolution in (\ref{ConvolutionTerm})
\begin{align}
(f*I)(\vec{x},t)&\longrightarrow \int_{\mathbb{R}^d} d^d x'f\left(\frac{\vec{x}}{\mu}-\vec{x}\,'\right)I(\vec{x}\,',t)\nonumber\\
&\hspace{1.6cm}=\int_{\mathbb{R}^d} d^d x'\frac{1}{\mu^d}\,f\left(\frac{\vec{x}-\vec{x}\,'}{\mu}\right)\,I(\vec{x}\,'/\mu,t)=:(f_\mu* I_\mu) (\vec{x},t)\,,\label{ReplaceConvolution}
\end{align}
where we have introduced the function
\begin{align}
f_\mu:\,\mathbb{R}^d&\longrightarrow \mathbb{R}_+\nonumber\\
\vec{x}&\longmapsto f_\mu(\vec{x})=\frac{1}{\mu^d}\,f(\vec{x}/\mu)\,.\label{DefIntermediateFunctions}
\end{align}

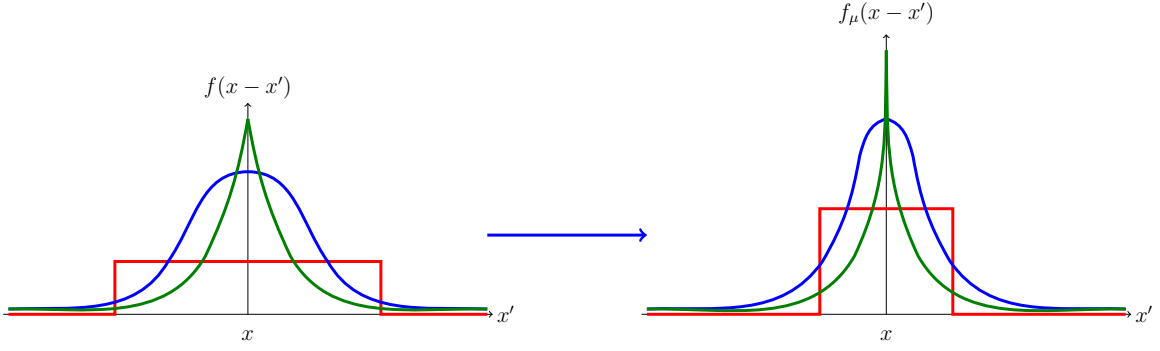
\begin{figure}
\begin{center}
\scalebox{0.7}{\parbox{22cm}{\begin{tikzpicture}
\draw[->] (-4.6,0) -- (4.6,0);
\node at (0,4.3) {$f(x-x')$};
\node at (4.85,0) {$x'$};
\node at (0,-0.4) {$x$};
\draw[->] (0,0) -- (0,4);
\draw[red,ultra thick] (-4.5,0)--(-2.5,0) -- (-2.5,1) -- (2.5,1) -- (2.5,0) -- (4.5,0);
\draw[blue,ultra thick] (0,2.7) to [out=0,in=125] (1.5,1) to [out=300,in=178] (4.5,0.1);
\draw[blue,ultra thick] (0,2.7) to [out=180,in=55] (-1.5,1) to [out=240,in=2] (-4.5,0.1);
\draw[green!50!black,ultra thick] (0,3.7) to [out=280,in=115] (0.8,1.1) to [out=300,in=179] (4.5,0.1);
\draw[green!50!black,ultra thick] (0,3.7) to [out=260,in=65] (-0.8,1.1) to [out=240,in=1] (-4.5,0.1);
\draw[ultra thick,->,blue] (4.5,1.5) -- (7.5,1.5);
\begin{scope}[xshift=12cm]
\draw[->] (-4.6,0) -- (4.6,0);
\node at (0,5.7) {$f_\mu(x-x')$};
\node at (4.85,0) {$x'$};
\node at (0,-0.4) {$x$};
\draw[->] (0,0) -- (0,5.3);
\draw[red,ultra thick] (-4.5,0)--(-1.25,0) -- (-1.25,2) -- (1.25,2) -- (1.25,0) -- (4.5,0);
\draw[blue,ultra thick] (0,3.7) to [out=-15,in=105] (0.5,3) to [out=-80,in=120] (1.2,1) to [out=305,in=178] (4.5,0.1);
\draw[blue,ultra thick] (0,3.7) to [out=195,in=75] (-0.5,3) to [out=-100,in=60] (-1.2,1) to [out=235,in=2] (-4.5,0.1);
\draw[green!50!black,ultra thick] (0,5) to [out=271,in=115] (0.6,1.1) to [out=300,in=179] (4.5,0.1);
\draw[green!50!black,ultra thick] (0,5) to [out=269,in=65] (-0.6,1.1) to [out=240,in=1] (-4.5,0.1);\end{scope}
\end{tikzpicture}
}}
\caption{Schematic examples of the rescaled functions $f_\mu$ for $0<\mu<1$ in $d=1$.}
\label{Fig:SchematicFunctionsRecale}
\end{center}
\end{figure}

\noindent
Schematic examples of the rescaled functions are shown in Figure~\ref{Fig:SchematicFunctionsRecale}: for $0<\mu<1$, the functions become more peaked around the origin, however, in such a way that the total area below the graph remains invariant. Indeed, for $f\in L^1(\mathbb{R}^d)$ (satisfying the first equation in (\ref{PropFuncF}) and $f(\vec{x})\geq 0$ $\forall \vec{x}\in\mathbb{R}^d$) and any $\mu>0$, the function $f_\mu$ has the following properties
\begin{enumerate}
\item[\emph{(i)}] $f_\mu\in L^1(\mathbb{R}^d)$, since 
\begin{align}
\int_{\mathbb{R}^d}d^dx\,|f_\mu(\vec{x})| =\frac{1}{\mu^d}\int_{\mathbb{R}^d}d^dx\,|f(\vec{x}/\mu)|=\int _{\mathbb{R}^d}d^d x\,|f(\vec{x})|<\infty\,,
\end{align}
where the last step follows due to $f\in L^1(\mathbb{R}^d)$.
\item[\emph{(ii)}] Due to the first equation in (\ref{PropFuncF})
\begin{align}
&\int_{\mathbb{R}^d}d^dx\,f_\mu(\vec{x})=\frac{1}{\mu^d}\int_{\mathbb{R}^d}d^dx\,f(\vec{x}/\mu)=\int _{\mathbb{R}^d}d^d x\,f(\vec{x})=\gamma\,.
\end{align}
\item[\emph{(iii)}] For any $\epsilon>0$
\begin{align}
\lim_{\mu\to 0}\int_{|\vec{x}|\geq \epsilon}\,|f_\mu(\vec{x})|=0\,,
\end{align}
which follows since
\begin{align}
\lim_{\mu\to 0}\int_{|\vec{x}|\geq \epsilon}d^dx\,|f_\mu(\vec{x})|=\lim_{\mu\to 0}\int_{|\vec{x}|\geq \epsilon}d^dx\,\frac{1}{\mu^d}\,|f(\vec{x}/\mu)|=\lim_{\mu\to 0}\int_{|\vec{x}|\geq \frac{\epsilon}{\mu}}d^d x\,f(\vec{x})\longrightarrow 0\,.
\end{align}
\end{enumerate}
As explained in Appendix~\ref{Sect:ApproxIdentity}, the family of functions $\{f_\mu\}_{\mu>0}$ is therefore an \emph{approximation to the identity}. 

The replacement (\ref{ReplaceConvolution}) leads to the following family of differential equations
\begin{align}
&\frac{\partial I_\mu}{\partial t}(\vec{x},t)=\left(1-I_\mu(\vec{x},t)\right)\,\left((f_\mu*I_\mu)(\vec{x},t)\right)\,,&&\forall \mu>0\,.\label{EqDiffSimpleEpidemicRenorm}
\end{align}
This is the same type of differential equation as in (\ref{EqDiffSimpleEpidemicRed}), except that the 'interaction' between infectious and susceptible individuals is now governed by a different function $f_\mu$. From an epidemiological perspective, we are interested in the continuum limit $\mu\to 0$. Using the properties of the family of functions $\{f_\mu\}_{\mu>0}$ and denoting $I_0(\vec{x},t)=\lim_{\mu\to \infty}I_\mu(\vec{x},t)$, we therefore find in the limit
\begin{align}
\lim_{\mu\to 0} (f_\mu*I_\mu)(\vec{x},t)=\gamma\,I_0(\vec{x},t)\,,
\end{align}
such that we find for the limit $\lim_{\mu\to 0}$ of the differential equation (\ref{EqDiffSimpleEpidemicRenorm})
\begin{align}
\frac{\partial I_0}{\partial t}(\vec{x},t)&=\gamma\,\left(1-I_0(\vec{x},t)\right)\,I_0(\vec{x},t)\,.\label{LimitingEquation}
\end{align}
We note here that despite the dependence of $I_0(\vec{x},t)$, the dynamics in the limit is entirely local at $\vec{x}$: the temporal evolution of the number of infectious individuals at $\vec{x}$ is independent of the number of infectious individuals at $\vec{x}\,'\neq\vec{x}$. From the perspective of the lattice, $\Delta X$ is now averaging the number of infectious over an infinite subset of the lattice $\Gamma$. From an epidemiological perspective, we are now averaging over a continuous function of infectious individuals in the geographic region $\Delta X$. If the latter represents \emph{e.g.} a country (labelled by $\vec{x}$), the temporal evolution of the disease no longer takes into account any geographic spread within the country. Furthermore, this evolution is completely independent of the evolution in neighbouring countries (\emph{i.e.} for $\vec{x}\,'\neq \vec{x}$) and describes the independent spread of a disease in different 'countries'. For given fixed $\vec{x}$ the equation (\ref{LimitingEquation}) is equivalent to the logistic equation stemming from the SI-model (without recovery) \cite{Kermack:1927}, which is among the simplest compartmental models (with only two compartments).

Going beyond the limit $\mu\to 0$ and following the example of the heat equation in Section~\ref{Sect:HeatEq}, we can consider an expansion of (\ref{EqDiffSimpleEpidemicRenorm}) for small $\mu$. To this end, we define
\begin{align}
I_\mu(\vec{x},t)=I^{(0)}(\vec{x},t)+\mu\,I^{(1)}(\vec{x},t)+\mathcal{O}(\mu^2)\,.
\end{align}
The main difficulty in expanding (\ref{EqDiffSimpleEpidemicRenorm}) is the convolution of the function $f_\mu$ with $I_\mu$ of the following form
\begin{align}
(f_\mu * I_\mu)(\vec{x},t)=\int_{\mathbb{R}^d}\frac{d^dx'}{\mu^d}\,f\left(\frac{\vec{x}-\vec{x}\,'}{\mu}\right)\left(I^{(0)}(\vec{x}\,',t)+\mu\,I^{(1)}(\vec{x}\,',t)+\mathcal{O}(\mu^2)\right)
\end{align}
Upon changing variables $\vec{y}=\frac{\vec{x}-\vec{x}\,'}{\mu}$ we can re-write this convolution in the form
\begin{align}
(f_\mu * I_\mu)(\vec{x},t)&=\int_{\mathbb{R}^d}d^dy\,f(\vec{y})\left(I^{(0)}(\vec{x}-\mu\,\vec{y},t)+\mu\,I^{(1)}(\vec{x}-\mu\,\vec{y},t)+\mathcal{O}(\mu^2)\right)\nonumber\\
&=\gamma\,I^{(0)}(\vec{x},t)+\mu\left[\gamma\,I^{(1)}(\vec{x},t)-\int_{\mathbb{R}^d}d^dy\,f(\vec{y})(\vec{y}\cdot \vec{\nabla}_x)I^{(0)}(\vec{x},t)\right]+\mathcal{O}(\mu^2)\,.\label{ConvolutionExpansion}
\end{align}
Equation (\ref{EqDiffSimpleEpidemicRenorm}) can then be expanded to first order to yield
\begin{align}
&\text{order }\mathcal{O}(\mu^0):&&\frac{\partial I^{(0)}}{\partial t}(\vec{x},t)=\gamma\,(1-I^{(0)}(\vec{x},t))\,I^{(0)}(\vec{x},t)\,,\label{LeadingOrderMollisonEq}\\
&\text{order }\mathcal{O}(\mu^1):&&\frac{\partial I^{(1)}}{\partial t}(\vec{x},t)=\gamma\,(1-2 I^{(0)}(\vec{x},t))\,I^{(1)}(\vec{x},t)\nonumber\\
& &&\hspace{2.4cm}-(1-I^{(0)}(\vec{x},t))\int_{\mathbb{R}^d}d^dy\,f(\vec{y})\,(\vec{y}\cdot \vec{\nabla}_x)\,I^{(0)}(\vec{x},t)\,.\label{FirstOrderMollisonEq}
\end{align}
The solution to (\ref{LeadingOrderMollisonEq}) is a logistic function (see (\ref{Background}) below), which may acquire an $\vec{x}$ dependence through the initial conditions. The solution $I^{(0)}$ enters into (\ref{FirstOrderMollisonEq}) as the coefficient of the term linear in $I^{(1)}$ as well as an inhomogeneity (which encodes additional information of the function $f$ beyond the simple constant $\gamma$). Notice, however, that the latter vanishes if initial conditions for $I^{(0)}$ are chosen that are independent of $\vec{x}$. 
\subsection{Length Scales at the Level of Solutions}
Formally, the family of equations (\ref{EqDiffSimpleEpidemicRenorm}) established in the previous Subsection can be thought of as being obtained from the simple epidemic model (\ref{EqDiffSimpleEpidemicRed}) through the transformation 
\begin{align}
&I\longrightarrow I_\mu\,,&&\text{and} &&f\longrightarrow f_\mu\,.\label{RenormalisationStep}
\end{align}
From a physics perspective this transformation leads to a family of effective models in which the 'interaction' of infectious and susceptible individuals is (through the function $f_\mu$) governed by the dimensionless scaling parameter $\mu$. In this Subsection we study the behaviour of particular classes of (approximate) solutions to the equations (\ref{EqDiffSimpleEpidemicRenorm}) and consider their behaviour under rescaling or the inclusion of a characteristic length scale.

\subsubsection{Travelling Waves and Similarity Transformation in $d=1$}\label{Sect:TravellingWaveSI}
\begin{figure}[htbp]
\begin{center}
\includegraphics[width=7.5cm]{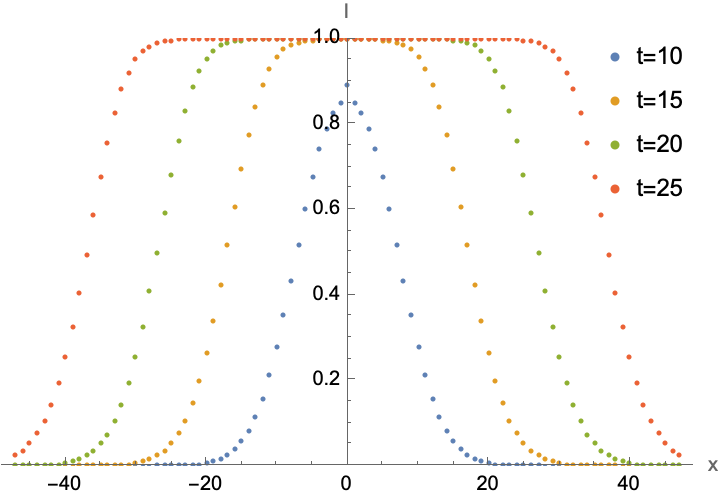}\hspace{1cm}\includegraphics[width=7.5cm]{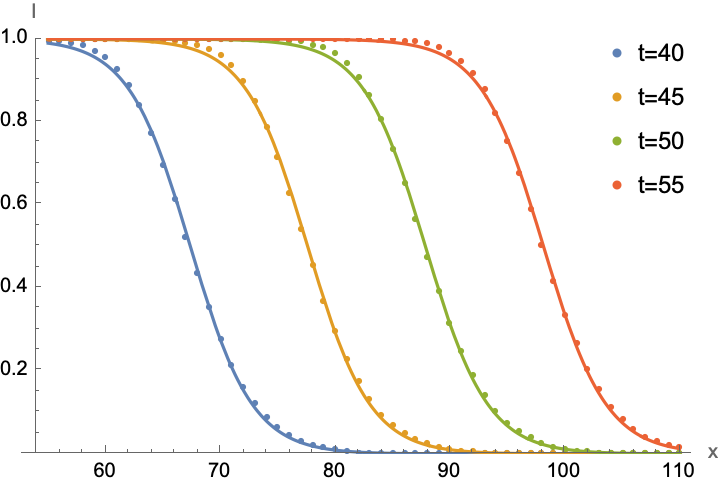}
\end{center}
\caption{Numerical solutions of (\ref{EqDiffSimpleEpidemic}) for $d=1$ and the initial conditions $I(X,t=0)=\delta_{X,0}/4$ for $F(X)=\frac{e^{-X^2/16}}{10}$ and $a=1$. Left panel: full solution shortly after the initial conditions. Right panel: the wave travelling to the right after sufficiently large time for the system to reach a state of intermediate asymptotics. The solid lines are fits of the numerical solution with a function of the form~(\ref{FitWaveTravel}).}
\label{Fig:TravellingWaves}
\end{figure}

As a first type of solutions, we shall consider travelling waves for $d=1$. Indeed, in one dimension and for the particular choice $f(\vec{x})=\frac{1}{2}\beta\,e^{-\beta|x|}$ (with $\beta\in\mathbb{R}_+$) travelling wave solutions of (\ref{EqDiffSimpleEpidemicRed}) have been found and discussed in \cite{MollisonVelocities} (see also \cite{KPP1937}) and it has been speculated that such solutions also exist for other classes of functions. A travelling wave is a solution $I_{w}:\,\mathbb{R}\rightarrow \mathbb{R}_+$ of (\ref{EqDiffSimpleEpidemicRed}) that depends on $(x,t)$ only through the particular combination $z:=x-vt$ for some constant velocity $v$. Inserting $I(x,t)=u(x-vt)$ into (\ref{EqDiffSimpleEpidemicRed}) 
\begin{align}
-v\,u'(z)=(1-u(z))\int_{-\infty}^\infty dx'\,f(x-x')\,u(x'-vt)\,,
\end{align}
and changing variables to $x'-vt=\xi$, travelling wave solutions are characterised by the integro-differential equation
\begin{align}
u'(z)=\frac{u(z)-1}{v}\int_{-\infty}^\infty d\xi\,f\left(z-\xi\right)\,u(\xi)=\frac{u(z)-1}{v}\,(f* u)(z)\,.
\end{align}
At least approximately we can demonstrate the existence of such waves by numerically solving the discretised equations (\ref{EqDiffSimpleEpidemic}). For simplicity\footnote{Other functions that are sufficiently peaked around $x=0$ yield very similar results.}, we consider (for $d=1$) $F$ to be a Gaussian function $F(X)=\frac{e^{-X^2/16}}{10}$. Numerical solutions for the initial condition $I(X,t=0)=\delta_{X,0}/100$ at different times $t$ (and for $a=1$ are shown in Figure~\ref{Fig:TravellingWaves}). Since $F$ is a symmetric function, we also have $I(X,t)=I(-X,t)$. After a certain initial time (as shown in the left panel of Figure~\ref{Fig:TravellingWaves}), the system reaches a state of the intermediate asymptotics in the sense of \cite{Barenblatt} (see Appendix~\ref{App:IntermediateAsymptotics}), the solution corresponds to a steady travelling wave moving from $X=0$ to the left and right: indeed, focusing on $X>0$, $I(X,t)$ can be well approximated by a sigmoid function of the form

\begin{figure}[htbp]
\begin{center}
\includegraphics[width=7.5cm]{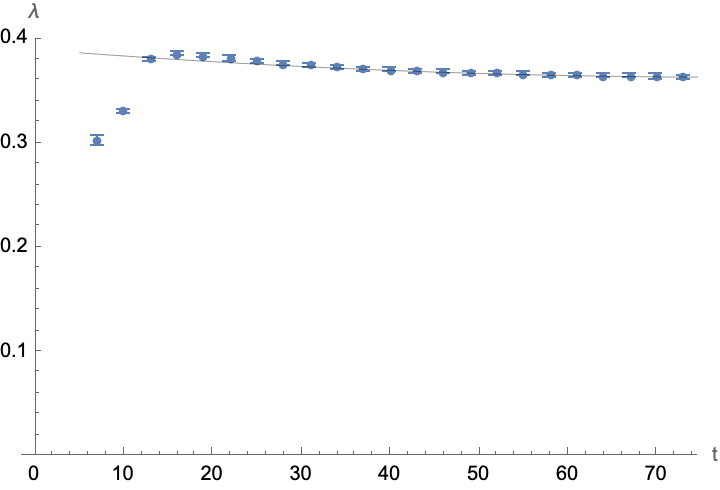}\hspace{1cm}\includegraphics[width=7.5cm]{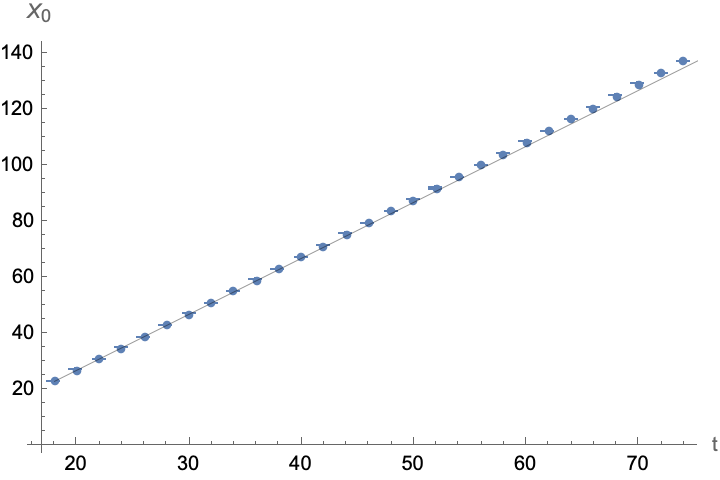}
\end{center}
\caption{Numerical plots of the functions $\lambda(t)$ and $X_0(t)$ as functions of $t$, along with the approximations in eq.~(\ref{ApproxPars})}
\label{Fig:GaussianParametersWave}
\end{figure}

\noindent
\begin{align}
&I(X,t)=1-\frac{1}{1+e^{-\lambda(t)(X-X_0(t))}}\,,&&\forall X>0\,.\label{IFitApprox}
\end{align}
This form is the same as (\ref{ApproxStochastic2}), which was used to approximate the stochastic model (\ref{StochasticEvolution}), thus justifying that (\ref{EqDiffSimpleEpidemic}) is a good (deterministic) approximation of the stochastic model (\ref{StochasticEvolution}). The quantities $\lambda(t)$ and $X_0(t)$ in (\ref{IFitApprox}) are functions of time, which are plotted in Figure~\ref{Fig:GaussianParametersWave}. These suggest that for large $t$ to good approximation\footnote{We note that here and in the following, the error estimates are only with regards to the fitting of the numerical solution.}

\noindent
\begin{align}
&\lambda(t)\sim\lambda_0\,,&&X_0(t)\sim\kappa+vt\,,&&\text{with} &&\begin{array}{l}\lambda_0=0.3900\pm 0.0008\,,\\ \kappa=-12.154\pm0.35\,,\\v=2.001\pm 0.002\,.\end{array}\label{ApproxPars}
\end{align}
Therefore, the solution $I(X,t)$ can (approximately) be written in the form
\begin{align}
I(X,t)=1-\frac{1}{1+e^{-\lambda_0(X-(\kappa+vt))}}=1-\frac{1}{1+B\,e^{-\lambda_0(X-vt)}}\,,&&\text{with} &&B=e^{\lambda_0\kappa}\,,\label{FitWaveTravel}
\end{align}
which is indeed only a function of the combination $X-vt$, where $v=$ const. and is therefore of the form of a steady travelling wave \cite{barenblatt_1996} of velocity $v$.

Next, we consider a rescaled model, in which we replace the function $F(X)$ by $F_\mu(X)$ following (\ref{Fig:SchematicFunctionsRecale})
\begin{align}
&F_\mu(X)=\frac{1}{\mu}\,F(X/\mu)=\frac{1}{10\,\mu}\,e^{-\frac{X^2}{16 \mu^2}}\,,&&\forall \mu>0\,.
\end{align}

\begin{figure}[htbp]
\begin{center}
\includegraphics[width=5cm]{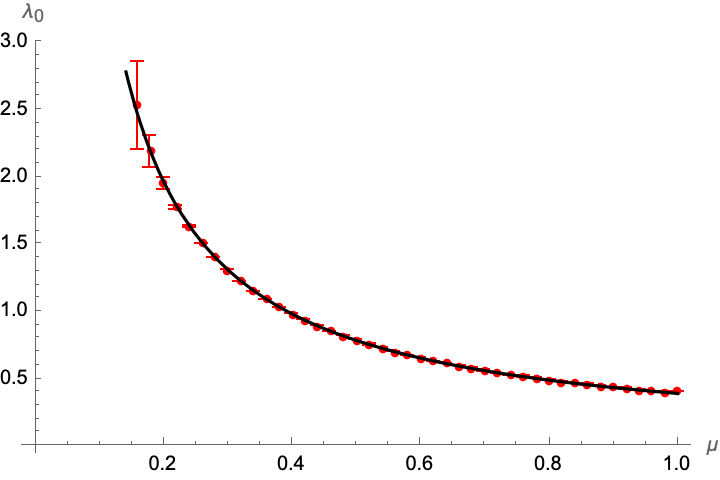}\hspace{0.5cm}\includegraphics[width=5cm]{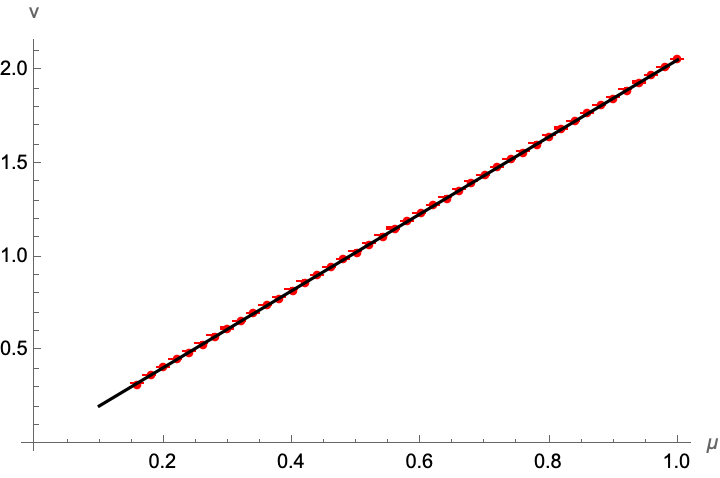}\hspace{0.5cm}\includegraphics[width=5cm]{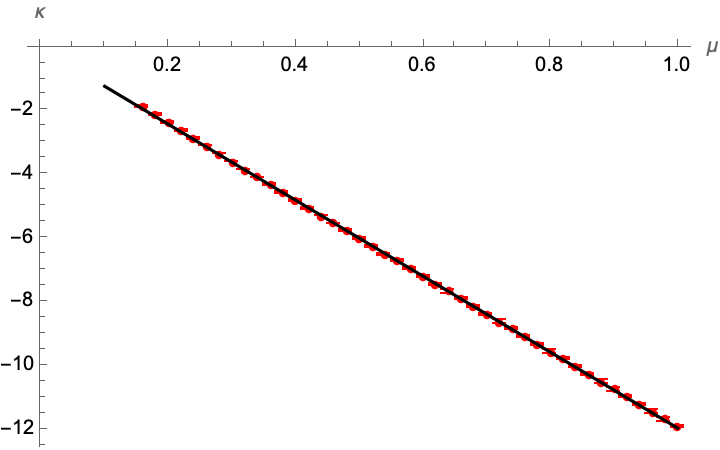}
\end{center}
\caption{Numerical plots of $\mu$ dependence of $(\lambda_0,v,\kappa)$ (red dots) and their approximation (solid black lines) using eq.(\ref{ApproxMuRescale})}
\label{Fig:RenormalisedParameters}
\end{figure}

\noindent
A numerical analysis suggest that the corresponding solutions $I_\mu(X,t)$ of (\ref{EqDiffSimpleEpidemic}) still correspond to traveling waves, which can be described by (\ref{FitWaveTravel}) except for coefficients $(\lambda_0,v,\kappa)$ that depend on $\mu$. Numerically, the functional dependence is plotted in Figure~\ref{Fig:RenormalisedParameters}, which can be fitted by
\begin{align}
\begin{array}{l}\lambda_0\sim \frac{\gamma_0}{\mu}\,,\\ v\sim \mu\, v_0\,,\\ \kappa\sim\mu\, \kappa_0\,,\end{array}&&\text{with} &&\begin{array}{l}\gamma_0=0.3941\pm 0.0002\,, \\ v_0=2.0581\pm0.0003\,, \\ \kappa_0=-11.93\pm0.01\,,  \end{array}\label{ApproxMuRescale}
\end{align}
such that we find for the solution
\begin{align}
I_\mu(X,t)=1-\frac{1}{1+e^{-\frac{\gamma_0}{\mu}(X-\mu(\kappa_0+v_0 t))}}=1-\frac{1}{1+B_\mu\,e^{-\gamma_0(X/\mu-v_0t)}}\,,&&\text{with} &&B_\mu=e^{\gamma_0\kappa_0}\,.
\end{align}
This is indeed approximately $I(X/\mu,t)$ with the function $I$ given in (\ref{FitWaveTravel}).\footnote{We note that a different choice of initial conditions also requires a shift $X\to X+c$ for some $c\in[-1,1]$. We furthermore note that our computation includes further sources of numerical errors (\emph{e.g.} the fact that the computations were performed on a finite array of $201$ points.)} This not only shows the consistency of the replacements (\ref{ReplaceFunc}) and (\ref{ReplaceConvolution}), it also shows that these latter can be seen as similarity transformations which have the structure of a group: performing two consecutive rescalings with parameters $\mu_1$ and $\mu_2$ is equivalent to performing a single rescaling with $\mu=\mu_1\mu_2$. Therefore (\ref{ReplaceFunc}) and (\ref{ReplaceConvolution}) bear the hallmark of a renormalisation group transformation.

Before continuing to other solutions we remark that the existence of travelling wave solutions of (\ref{EqDiffSimpleEpidemicRenorm}) for $d>1$ is a more delicate question. Indeed, substitution of the ansatz $I_\mu(\vec{x},t)=u(z)$ with $z=|\vec{x}|-vt$ leads to 
\begin{align}
-v\,u'(z)=(1-u)\int_{\mathbb{R}^d}d^dx'f_\mu(\vec{x}-\vec{x}\,')\,u(|\vec{x}\,'|-vt)\,.
\end{align}
For general $L^1(\mathbb{R}^d)$-functions, the latter, however, is not consistent, since the right hand side is not necessarily a function of $z$ alone.

\subsubsection{Linearisation Around a Position Independent Background}\label{Sect:Linearisation}
As another type of solution of (\ref{EqDiffSimpleEpidemicRed}) we consider a linearisation around a position independent solution
\begin{align}
&I(\vec{x},t)=I_0(t)+h(\vec{x},t)\,,&&\text{with} &&\begin{array}{l}I_0:\,\mathbb{R}\longrightarrow \mathbb{R}_+\,,\\h(\vec{x},t):\, \mathbb{R}^d\times \mathbb{R}\longrightarrow\mathbb{R}\,.\end{array}\label{LinearisingSolution}
\end{align}
where we consider $h$ to be a small fluctuation. Expanding (\ref{EqDiffSimpleEpidemicRed}) for small $h$ we find to leading order (\ref{LimitingEquation}) \emph{i.e.}
\begin{align}
&\frac{\partial I_0}{\partial t}(t)=\gamma\,(1-I_0(t))\,I_0(t)&&\text{ with solution }&&I_0(t)=\frac{1}{1+e^{-\gamma(t-\tau_0)}}\,,\label{Background}
\end{align}
with $\gamma$ given in (\ref{PropFuncF}) and $\tau_0\in\mathbb{R}_+$ is an integration constant. To first order in $h$, (\ref{EqDiffSimpleEpidemicRed}) becomes
\begin{align}
\frac{\partial h}{\partial t}(\vec{x},t)=-\gamma\,h(\vec{x},t)\,I_0(t)+(1-I_0(t))\,(f*h)(\vec{x},t)\,.\label{FirstOrder}
\end{align}
In (\ref{ConvolutionExpansion}) we have expanded the convolution $(f* h)$ for small scaling parameter $\mu$. Here we shall take a slightly different route and assume that for both $f$ and $h$ their Fourier transforms $\widehat{f}$ and $\widehat{h}$ exist and that the inversion can be written in the form
\begin{align}
&f(\vec{x})=\frac{1}{(2\pi)^{d/2}}\int_{\mathbb{R}^d}d^dk\,e^{i\vec{k}\cdot \vec{x}}\,\widehat{f}(\vec{k})\,,&&h(\vec{x},t)=\frac{1}{(2\pi)^{d/2}}\int_{\mathbb{R}^d}d^dk\,e^{i\vec{k}\cdot \vec{x}}\,\widehat{h}(\vec{k},t)\,.
\end{align}
The convolution of the two functions is then written as
\begin{align}
(f*h)(\vec{x},t)=\int_{\mathbb{R}} d^dk\,e^{i\vec{k}\cdot\vec{x}}\,\widehat{f}(\vec{k})\,\widehat{h}(\vec{k},t)\,,
\end{align}
such that (\ref{FirstOrder}) becomes
\begin{align}
\frac{1}{(2\pi)^{d/2}}\int_{\mathbb{R}^d}d^dk\,e^{i\vec{k}\cdot\vec{x}}\,\frac{\partial\widehat{h}}{\partial t}(\vec{k},t)=-\frac{\gamma I_0(t)}{(2\pi)^{d/2}}\,\int_{\mathbb{R}^d}d^dk\,e^{i\vec{k}\cdot\vec{x}}\,\widehat{h}(\vec{k},t)+(1-I_0(t))\int_{\mathbb{R}} d^dk\,e^{i\vec{k}\cdot\vec{x}}\,\widehat{f}(\vec{k})\,\widehat{h}(\vec{k},t)\,.\nonumber
\end{align}
We thus obtain the following differential equation for the Fourier transformation of $h$
\begin{align}
\frac{\partial\widehat{h}}{\partial t}(\vec{k},t)=-\gamma\,I_0(t)\,\widehat{h}(\vec{k},t)+(2\pi)^{d/2}\,(1-I_0(t))\,\widehat{f}(\vec{k})\,\widehat{h}(\vec{k},t)\,,
\end{align}
which can be integrated to yield
\begin{align}
\widehat{h}(\vec{k},t)=\widehat{h}(\vec{k},t_0)\,\text{exp}\left[\int_{t_0}^tdt' \left((2\pi)^{d/2}\,(1-I_0(t'))\,\widehat{f}(\vec{k})-\gamma\,I_0(t')\right)\right]\,.
\end{align}
To leading order we therefore find for $h(\vec{x},t)$
\begin{align}
h(\vec{x},t)=\frac{e^{-\gamma\int_{t_0}^t dt' I_0(t')}}{(2\pi)^{d/2}}\,\int_{\mathbb{R}^d}d^dk\, \widehat{h}(\vec{k},t_0)\,\text{exp}\left[(2\pi)^{d/2}\int_{t_0}^tdt' (1-I_0(t'))\,\widehat{f}(\vec{k})\right]\,e^{i\vec{k}\cdot\vec{x}}\,.\label{FormFluctSol}
\end{align}
To continue further, we next consider averaging the solution (\ref{LinearisingSolution}) over an interval $[-L,L]^{d}$ (with $L\in\mathbb{R}_+$), \emph{i.e.} we define
\begin{align}
\overline{I}_L(\vec{x},t)=\frac{1}{(2L)^d}\left(\prod_{i=1}^d \int_{-L+x_i}^{L+x_i}dx_i'\right)I(\vec{x}\,',t)=I_0(t)+\overline{h}_L(\vec{x},t)\,.\label{DefAveragedBackground}
\end{align}
Assuming (absolute) convergence of the integral in (\ref{FormFluctSol}), such that the order of integration of $\vec{k}$ and $\vec{x}\,'$ in (\ref{DefAveragedBackground}) can be exchanged, we can write explicitly (with $\vec{k}=(k_1,\ldots,k_d)$)
\begin{align}
\overline{h}_L(\vec{x},t)=\frac{e^{-\gamma\int_{t_0}^t dt' I_0(t')}}{(2\pi)^{d/2}}\,\int_{\mathbb{R}^d}d^dk\, \widehat{h}(\vec{k},t_0)\,\left(\prod_{i=1}^d\frac{\sin(k_i L)}{k_i L}\right)\,\text{exp}\left[(2\pi)^{d/2}\int_{t_0}^tdt' (1-I_0(t'))\,\widehat{f}(\vec{k})\right]\,e^{i\vec{k}\cdot\vec{x}}\,.\label{DefAveragedSolution}
\end{align}
The main difference of this expression compared to (\ref{FormFluctSol}) is that individual Fourier modes (characterised by $\vec{k}$) are weighted by the $L$-dependent factor $\prod_{i=1}^d\frac{\sin(k_i L)}{k_i L}$. This can lead to resonance effects: indeed, if for example the initial conditions are characterised by a dominant Fourier mode, specific choices of $L$ can lead to a strong suppression of the fluctuation $\overline{h}_L$, such that the averaged number of infected individuals is very well approximated by the (homogeneous) $I_0(t)$, which is the solution of the limiting equation (\ref{LimitingEquation}). The existence of a dominant Fourier mode in the initial conditions $h(\vec{x},t_0)$ indicates the presence of a certain length scale in the distribution of initial infectious individuals, which can for example be caused by the characteristic length distance between population centers and agglomerations. Such information is very useful to extract critical length scales for effective theories.

To further illustrate this idea, we shall discuss a concrete examples for $d=1$, with $f$ a Gaussian function  of the form
\begin{align}
&f(x)=f_0\,e^{-\beta x^2}\,,&&\text{with} &&f_0,\beta\in\mathbb{R}_+\,,
\end{align}
such that we have
\begin{align}
&\gamma=f_0\,\sqrt{\tfrac{\pi}{\beta}}\,, &&\text{and} && \widehat{f}(k)=\frac{f_0}{\sqrt{2\beta}}\,e^{-\frac{k^2}{4\beta}}\,.\label{AdditionalParameters}
\end{align}
For the initial conditions at $t=0$ we choose $I(x,t=0)=A\,(\cos(\alpha x)+1)$ (with $A,\alpha\in\mathbb{R}_+$). 
Decomposing these initial conditions according to (\ref{LinearisingSolution}) we have
\begin{align}
&\begin{array}{l}I_0(0)=A\,,\\h(x,0)=A\,\cos(\alpha x)\,,\end{array}&&\text{with}&&\widehat{h}(k,0)=\frac{A}{2}\,\sqrt{2\pi}\,\left(\delta(k-\alpha)+\delta(k+\alpha)\right)\,,
\end{align}
which leads to the background solution (\ref{Background}) $I_0(t)=\frac{A}{A+(1-A)\,e^{-\gamma t}}$
and the fluctuation (\ref{FormFluctSol})
\begin{align}
h(x,t)
&=\frac{A\,\text{exp}\left(\gamma\, t\, e^{-\frac{\alpha^2}{4\beta}}\right)}{1+A(e^{t\gamma}-1)}\,\cos(\alpha x)\,\left(1+A(e^{t\gamma}-1)\right)^{-\text{exp}\left(-\frac{\alpha^2}{4\beta}\right)}\,.
\end{align}
Given this explicit form of the fluctuation, we can verify whether it is indeed a small fluctuation compared to the homogeneous background solution (\ref{Background}). The function $h(x,t)$ satisfies
\begin{align}
&h(x,t=0)=A\,,&&\text{and} &&\lim_{t\to\infty} h(x,t)=0\,,
\end{align}
and has an extremum at
\begin{align}
t_{\text{ext}}=\frac{8\beta \text{arctanh}(1-2A)-\alpha^2}{4\beta\gamma}\,,
\end{align}
for which $I_0(t_{\text{ext}})=\frac{1}{1+e^{\alpha^2/(4\beta)}}$ which is independent of $A$, while $\lim_{A\to 0}h(x,t_{\text{ext}})=0$. Thus the fluctuation can be made arbitrarily small compared to the background solution if the initial fluctuation is sufficiently small. 

\begin{figure}[htbp]
\begin{center}
\includegraphics[width=7.5cm]{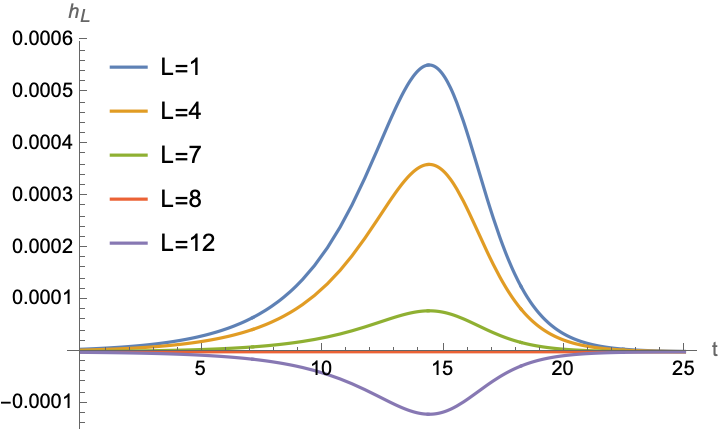}\hspace{1cm}\includegraphics[width=7.5cm]{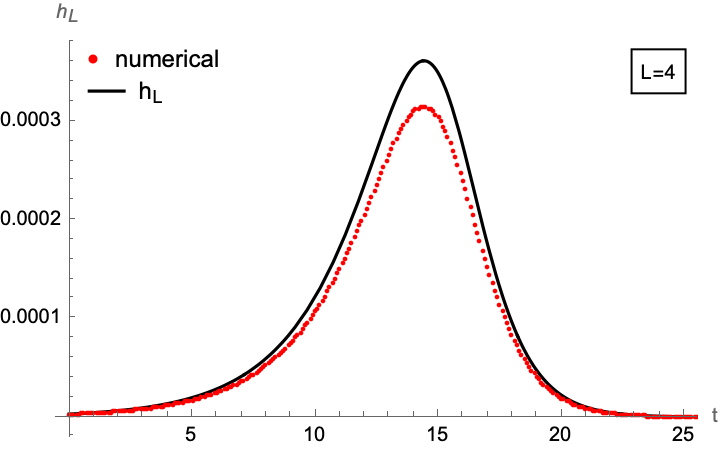}
\end{center}
\caption{Left panel: plots of $\overline{\mathfrak{h}}_L$ in (\ref{FluctuationDominantMode}) for different values of $L$. Right panel: comparison between $\overline{\mathfrak{h}}_L$ in (\ref{FluctuationDominantMode}) (solid black line) and a numerical solution of (\ref{EqDiffSimpleEpidemicRed}) (red) for $L=4$. Both plots use $A=1/200000$, $\alpha=\pi/8$, $f_0=1/10$ and $\beta=1/20$.}
\label{Fig:FluctuationLinear}
\end{figure}

The averaged fluctuation (\ref{DefAveragedSolution}) then takes the form
\begin{align}
\overline{h}_L(x,t)=\cos(\alpha x)\,\overline{\mathfrak{h}}_L(t)\,,&&\text{with}&&\overline{\mathfrak{h}}_L(t)=\frac{A\,\text{exp}\left(\gamma\, t\, e^{-\frac{\alpha^2}{4\beta}}\right)}{1+A(e^{t\gamma}-1)}\,\frac{\sin(L\alpha)}{L\alpha}\,\left(1+A(e^{t\gamma}-1)\right)^{-\text{exp}\left(-\frac{\alpha^2}{4\beta}\right)}\,.\label{FluctuationDominantMode}
\end{align}
Graphical plots of $\overline{\mathfrak{h}}_L$ for different values of $L$, as well as a comparison to the numerical solution of the full non-linear equation (\ref{EqDiffSimpleEpidemicRed}) are shown in Figure~\ref{Fig:FluctuationLinear}. As is evident from (\ref{FluctuationDominantMode}), the fluctuation $\overline{\mathfrak{h}}_L$ is suppressed by an inverse power of $L$. Moreover, if $L\alpha$ is an integer multiple of $\pi$, the fluctuation is vanishing altogether and the number of infected individuals is very well approximated by $I_0$, which is a solution of the limiting equation (\ref{LimitingEquation}).

This fact can also be visualised in a different fashion: averaged solutions of (\ref{EqDiffSimpleEpidemicRed}) can be roughly approximated by a logistic function 
\begin{align}
\overline{I}_L(t)=\frac{1}{2L} \int_{-L}^L I(x,t)\simeq \frac{1}{1+e^{-\kappa(t-t_0)}}\,.\label{EffectiveLogistic}
\end{align}

\begin{figure}[htbp]
\begin{center}
\includegraphics[width=7.5cm]{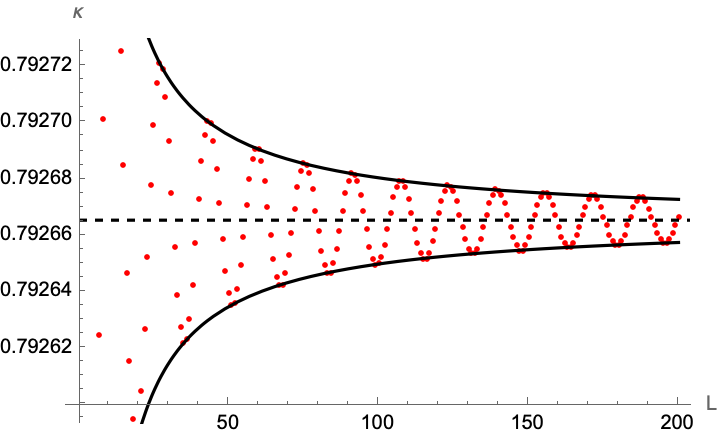}\hspace{1cm}\includegraphics[width=7.5cm]{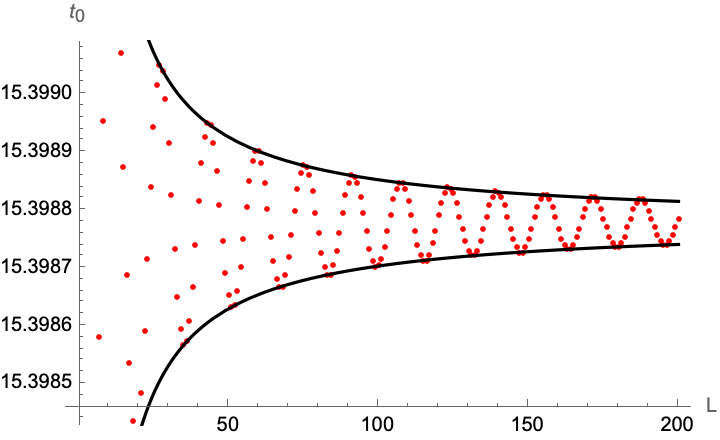}
\end{center}
\caption{Numerical fits of $\kappa$ (left) and $t_0$ (right) in (\ref{EffectiveLogistic}). Both plots use $A=1/200000$, $\alpha=\pi/8$, $f_0=1/10$ and $\beta=1/20$. The dashed horizontal line in the left plot represents the value of $\gamma$ in (\ref{AdditionalParameters}). The enveloping curves in black are compatible with a $1/L$ behaviour. }
\label{Fig:EffectiveParameters}
\end{figure}

\noindent
Numerical fits for the effective values of $\kappa$ and $t_0$ as functions of $L$ are plotted in Figure~\ref{Fig:EffectiveParameters}. The effective description oscillates around the solution of the limiting equation (\ref{LimitingEquation}), with an amplitude that is roughly decaying as $1/L$.
\section{Conclusions and Outlook}\label{Sect:Conclusions}
In this paper we have studied effective models and renormalisation group methods for simple epidemiological theories, following the example of diffusion processes, as is schematically shown in the following diagram:

\begin{center}
\scalebox{1}{\parbox{16.5cm}{\begin{tikzpicture}
\node at (5.5,-0.85) {{\bf diffusion}};
\node at (12,-0.85) {{\bf epidemiological model}};
\draw[fill=red!60!white] (-1.9,-1.75) -- (1.95,-1.75) -- (1.95,-2.3) -- (-1.9,-2.3) -- (-1.9,-1.75);
\node at (0,-2) {{\bf stochastic model}};
\node at (5.5,-2) {\parbox{2.75cm}{\small random walks\\probability (\ref{PFourierInt})}};
\node at (12,-2) {\parbox{3.9cm}{\small stochastic lattice model\\probability (\ref{StochasticEvolution})}};
\draw[red,ultra thick,->] (0,-2.3) -- (0,-4.2);
\node[red,rotate=90] at (-0.2,-3.2) {\tiny approximation};
\node[red,rotate=90] at (0.2,-3.2) {\tiny coarse graining};
\draw[fill=blue!50!white] (-1.9,-4.25) -- (1.95,-4.25) -- (1.95,-5.4) -- (-1.9,-5.4) -- (-1.9,-4.25);
\node at (0,-4.8) {\parbox{3.5cm}{\centering {\bf deterministic effective model}}};
\node at (5.5,-4.8) {\parbox{6cm}{\small deformed heat eq.~(\ref{EqExpanddiff2})\\deformed probability density (\ref{paExpansion})}};
\node at (12,-4.8) {\parbox{3.9cm}{\centering\small 'simple epidemic model'~ eq.(\ref{EqDiffSimpleEpidemicRenorm})}};
\draw[red,ultra thick,->] (0,-5.4) -- (0,-6.9);
\node[red,rotate=90] at (-0.2,-6.1) {\tiny continuum};
\node[red,rotate=90] at (0.2,-6.1) {\tiny limit};
\draw[fill=green!50!white] (-1.9,-6.95) -- (1.95,-6.95) -- (1.95,-7.65) -- (-1.9,-7.65) -- (-1.9,-6.95);
\node at (0,-7.3) {\parbox{3.5cm}{\centering {\bf limiting model}}};
\node at (5.5,-7.3) {\parbox{3.9cm}{\small heat equation (\ref{DefHeatEq})\\Green's function (\ref{CausalGreensFunction})}};
\node at (12,-7.3) {\parbox{3.9cm}{\centering\small SI-model~ (\ref{LimitingEquation})}};
\end{tikzpicture}
}}
\end{center}

\noindent
The starting point is a stochastic approach, which models the microscopic dynamics (\emph{i.e.} the infection among individuals) in a probabilistic manner. Such models can be approximated by deterministic models, which also include a step of coarse graining, \emph{i.e.} discrete time- and space-variables are replaced by constant ones, by averaging over suitable regions of the coordinate space. At this point we assume that the variables do not vary too much over the averaged region, which corresponds to a status of \emph{intermediate asymptotic} in the language of \cite{IntermediateAsymptotics,Barenblatt} (see Appendix~\ref{App:IntermediateAsymptotics}). Rescaling the spatial variable in this model leads to a family of effective models, which are distinguished by the size of the population, over which an averaging has taken place. In this way, the initial microscopic details of the epidemiological process become less and less important: indeed, while the latter in the case of the 'simple epidemic model' \cite{Mollison1977,MollisonVelocities} are encoded in an $L^1$-function $f$, in the continuum limit only its $L^1$-norm remains, which is interpreted as the infection rate of the resulting SI-model. Mathematically, this is due to the fact that the series of functions (called $f_\mu$ in (\ref{DefIntermediateFunctions})) that govern the infection in an arbitrary member of the family of effective models, form an approximation to the identity (see Appendix~\ref{Sect:ApproxIdentity}). In Section~\ref{Sect:SimpleEpidemicSI}, we have studied the transition of the effective models towards the continuum limit, both at the level of the defining differential equation and its solutions. Since the transition from one effective model to another takes the form of a self-similarity transformation, we interpret it as a renormalisation group transformation, in the sense of~\cite{ChenGoldenfeldOono1995} .

While the type of models we discuss in this work are simplistic from an epidemiological perspective, they still allow to draw some conceptual conclusions: in order to describe the epidemiological evolution on a small population (\emph{i.e.} for $\mu\gg 0$), the detailed structure of the function $f_\mu$ in (\ref{DefIntermediateFunctions}) is crucial. In other words, to describe the spread of the disease in a small group of people, the (microscopic) details of how the disease is passed on are important to drive the dynamics. For larger groups, these details become less important, while in the continuum limit only a single number (the infection rate) remains relevant. This explains, why simple epidemiological descriptions, such as the eRG \cite{DellaMorte:2020wlc,DellaMorte:2020qry,cacciapaglia2020second,cacciapaglia2020mining,cacciapaglia2020evidence,cacciapaglia2020multiwave,Cacciapaglia:2020mjf,cacciapaglia2020us,Cacciapaglia:2021cjl,
Cacciapaglia:2021vvu,GreenPass,MLvariants} are capable of successfully describing large scale developments (\emph{e.g.} the time evolution at the level of countries). We have furthermore seen indications in Section~\ref{Sect:Linearisation} that the crucial length scale above which such effective models can be used, may also depend on the initial conditions of the problem and require a more careful analysis of the solutions.

In the the future, we plan to extend our results to include more realistic features into the models, such as the possibility of recovery, non-instantaneous infection among individuals, the mobility of the population or the possibility of the spread of competing variants of a single disease.

\section*{Acknowledgements}
We thank E.~Battista, G.~Cacciapaglia, C.~Cot, D.~Iacobacci and S.Vatani for enlightening discussions and exchanges. This study is supported by Ministero dell’Università e della Ricerca (Italy), Piano Nazionale di Ripresa e Resilienza, and EU within the Extended Partnership initiative on Emerging Infectious Diseases project number PE00000007 (One Health Basic and Translational Actions Addressing Unmet Needs on Emerging Infectious Diseases). The work of F.S. is partially supported by the Carlsberg Foundation, semper ardens grant CF22-0922.

\appendix
\section{Notation and Conventions}
\subsection{Approximation of the Identity}\label{Sect:ApproxIdentity}
{\bf Def:} (see \emph{e.g.} \cite{SteinWeiss,Rudin,Folland,Grafakos}) An {\bf approximation of the identity} (also called \emph{approximate identity}) is a family of functions $\{g_\mu\}_{\mu>0}$ with $g_\mu \in L^1(\mathbb{R}^d)$ such that   
\begin{enumerate}
\item $\{g_\mu\}$ is bounded in the $L^1(\mathbb{R}^d)$ norm, \emph{i.e.} $\exists\, c>0$ such that $||g_\mu||\leq c$ $\forall \mu>0$
\item $\int_{\mathbb{R}^d}g_\mu (x)\,dx=1$ $\forall \mu>0$
\item For any $\epsilon>0$: $\lim_{\mu\to 0}\int_{|\vec{x}|\geq \epsilon}|g_\mu(x)|dx=0$.
\end{enumerate}

\noindent
Let $\{g_\mu\}_{\mu>0}$ be an approximation to the identity, we then have
\begin{enumerate}
\item If $f\in L^p(\mathbb{R}^d)$ (with $1\leq p<\infty$), then $f*g_\mu\in L^p(\mathbb{R}^d)$ and $f* g_\mu\longrightarrow f$ in the $L^p(\mathbb{R}^d)$ norm, as $\mu\to 0$.
\item If $f$ is uniformly continuous and bounded on $\mathbb{R}^d$, then $f*g_\mu$ is also uniformly continuous and bounded on $\mathbb{R}^d$. Furthermore, $f*g_\mu\longrightarrow f$ uniformly, \emph{i.e.} in the $L^\infty(\mathbb{R}^d)$ norm, as $\mu\to 0$.
\item If $f\in L^\infty(\mathbb{R}^d)$ is continuous on an open subset $U\subset\mathbb{R}^d$, then $f*g_\mu\in L^\infty (\mathbb{R}^d)$, is uniformly continuous on $\mathbb{R}^d$ and $f*g_\mu\longrightarrow f$ uniformly on compact subsets $K\subset U$, as $\mu\to 0$.
\end{enumerate} 
\subsection{Review of Intermediate Asymptotics and Renormalisation Group}\label{App:IntermediateAsymptotics}
Our main reference for this appendix is \cite{IntermediateAsymptotics}, building on the earlier work \cite{Barenblatt}: consider a dimensionless quantity $Z$ that depends on $n+1$ dimensionless variables $\{\mu,z_1,\ldots,z_n\}$
\begin{align}
Z=f(\mu,z_1,\ldots,z_n)\,.
\end{align}
Following \cite{Barenblatt}, the system is said to be in an \emph{intermediate asymptotic} state, if its dynamics is independent on the exact initial and/or boundary conditions, but has not reached a final (\emph{i.e.} static) state. If $\mu\to 0$ corresponds to an intermediate asymptotic limit, there are three different possibilities \cite{Barenblatt} (see also \cite{IntermediateAsymptotics})
\begin{itemize}
\item \emph{self-similarity of the first kind:} the limit $\lim_{\mu} f(\mu,z_1,\ldots,z_n)$ exists
\item \emph{self-similarity of the second kind:} the limit $\lim_{\mu} f(\mu,z_1,\ldots,z_n)$ is not well defined, but instead there exist $(\alpha,\alpha_1,\ldots,\alpha_n)\in\mathbb{R}^{n+1}$ such that the limit
\begin{align}
\lim_{\mu\to 0} \frac{Z}{\mu^\alpha}=\lim_{\mu\to 0} \frac{1}{\mu^\alpha}\,f\left(\mu,\frac{z_1}{\mu^{\alpha_1}}\,,\ldots\,,\frac{z_n}{\mu^{\alpha_n}}\right)\,,\label{IAmultiscaling}
\end{align}
exists. The parameters $(\alpha,\alpha_1,\ldots,\alpha_n)\in\mathbb{R}^{n+1}$ cannot be obtained by dimensional analysis alone, but they following in principle from the differential equation that describes the dynamics of $Z$. As remarked in \cite{IntermediateAsymptotics}, from the perspective of the RG, the parameters $(\alpha,\alpha_1,\ldots,\alpha_n)\in\mathbb{R}^{n+1}$ are the anomalous dimensions.
\item None of the above, in which case a different rescaling than (\ref{IAmultiscaling}) might be required.
\end{itemize}
\section{Mathematical Expansion}\label{App:ErfExpansion}
In this appendix we calculate $p_a$ in (\ref{paExpansion}) up to order $\mathcal{O}(a^2)$. To this end, we begin by writing
\begin{align}
p_a(\vec{x},t)&=\frac{1}{(2\pi)^d}\prod_{i=1}^d\int_{\left[-\frac{\pi}{a},\frac{\pi}{a}\right]} dk_i\,e^{i k_i x_i}\left[1-\frac{a^2}{2d}\sum_{j=1}^d k_j^2+\frac{a^4}{24 d}\sum_{j=1}^dk_j^4+\mathcal{O}(a^6)\right]^{\frac{2d\kappa t}{a^2}}\nonumber\\
&=\frac{1}{(2\pi)^d}\prod_{i=1}^d\int_{\left[-\frac{\pi}{a},\frac{\pi}{a}\right]} dk_i\,e^{i k_i x_i}\,e^{-\kappa t\sum_{j=1}^dk_j^2}\left[1-\frac{\kappa t a^2}{12d}\left(3\left(\sum_{j=1}^d k_j^d\right)^2-d\sum_{j=1}^d k_j^4\right)\right]+\mathcal{O}(a^4)\,.\label{ExpansionOrdera}
\end{align}
The first term in the square bracket yields the limiting contribution (\ref{CausalGreensFunction}) up to exponentially suppressed terms (stemming from the dependence of the integration boundaries on $a$), which we are not interested in. Similarly, for the second term in the square bracket in (\ref{ExpansionOrdera}), the dependence of the integral boundaries on $a$ only produce exponentially suppressed contributions, which we neglect. To order $\mathcal{O}(a^2)$ we can therefore write
\begin{align}
p_a(\vec{x},t)
=p(\vec{x},t)-\frac{a^2 \kappa t}{12 d(2\pi)^2}\int_{\mathbb{R}^d}d^dk\,e^{-\sum_{j=1}^d(\kappa t k_j^2-ik_j x_j)}\,\left[(3-d)\sum_{j=1}^d k_j^4+6\sum_{1\leq i<j\leq d}k_i^2 k_j^2\right]+\mathcal{O}(a^4)\,.
\end{align}
The remaining integrals are of Gauss type, for which we can use the results
{\allowdisplaybreaks
\begin{align}
\int_{\mathbb{R}}dk_j \,e^{ik_j x_j-\kappa t k_j^2}&=\sqrt{\frac{\pi}{\kappa t}}\,e^{-\frac{x_j^2}{4\kappa t}}\,,\nonumber\\
\int_{\mathbb{R}}dk_j \,k_j^2\,e^{ik_j x_j-\kappa t k_j^2}&=\frac{(2\kappa t-x_j^2)\sqrt{\pi}}{4(\kappa t)^{5/2}}\,e^{-\frac{x_j^2}{4\kappa t}}\,,\nonumber\\
\int_{\mathbb{R}}dk_j \,k_j^4\,e^{ik_j x_j-\kappa t k_j^2}&=\frac{(12\kappa^2 t^2-12 \kappa t x_j^2+x_j^4-x_j^2)\sqrt{\pi}}{16(\kappa t)^{9/2}}\,e^{-\frac{x_j^2}{4\kappa t}}\,,
\end{align}}
such that we find
\begin{align}
p_a(\vec{x},t)&=p(\vec{x},t)-\frac{a^2 \pi^{d/2} e^{-\frac{|\vec{x}|^2}{4\kappa t}}}{192d (2\pi)^d (\kappa t)^{d/2+3}}\,\bigg[(3-d)\sum_{j=1}^d\left(x_j^4-12\kappa t x_j^2+12\kappa^2 t^2\right)\nonumber\\
&\hspace{4.5cm}+6\sum_{1\leq i<j\leq d}\left(x_i^2x_j^2-2\kappa t (x_i^2+x_j^2)+4\kappa^2t^2\right)\bigg]+\mathcal{O}(a^4)\,.
\end{align}
Using the form of (\ref{CausalGreensFunction}), this result can be simplified to give
\begin{align}
p_a(\vec{x},t)&=\frac{e^{-\frac{|\vec{x}|^2}{4\kappa t}}}{(4\pi \kappa t)^{d/2}}\left[1-\frac{a^2}{192d(\kappa t)^3}\left(3|\vec{x}|^4-d\sum_{j=1}^d x_j^4-24 \kappa t |\vec{x}|^2+24 d \kappa^2 t^2\right)\right]+\mathcal{O}(a^4)\,.\label{PerturbGreen}
\end{align}


\bibliographystyle{ieeetr}
\bibliography{biblio}

\end{document}